\documentclass[prb,aps,floatfix,floats,superscriptaddress]{revtex4}
\usepackage{epsfig}
\usepackage{color}
\usepackage{amsmath}
\usepackage{amssymb}
\newcommand{\be}{\begin{equation}}
\newcommand{\ee}{\end{equation}}
\newcommand{\bea}{\begin{eqnarray}}
\newcommand{\eea}{\end{eqnarray}}

\def\fa#1#2{{a_#1(#2)}}
\def\lp{{\Lambda^\prime}}

\def\da{\downarrow}
\def\fnn{\nonumber\\}
\def\l{\Lambda}
\def\a{\alpha}
\def\b{\beta}
\def\g{\gamma}

\def\th{\theta}
\def\nn{\nonumber\\}

\def\fr#1{(\ref{#1})}

\def\eps{\epsilon}

\def\K{{\cal K}}
\begin{document}
\title{Threshold Singularities in the One Dimensional Hubbard Model}
\author{Fabian H.L. Essler}
\affiliation{
The Rudolf Peierls Centre for Theoretical Physics, Oxford
University, Oxford OX1 3NP, UK}
\begin{abstract}
We consider excitations with the quantum numbers of a hole in the one
dimensional Hubbard model below half-filling. We calculate the
finite-size corrections to the energy. The results are then used to
determine threshold singularities in the single-particle Green's
function for commensurate fillings. We present the analogous results
for the Yang-Gaudin model (electron gas with $\delta$-function
interactions). 
\end{abstract}
\maketitle
\section{Introduction}
The Hubbard model constitutes a key paradigm for strong correlation
effects in one dimensional (1D) electron systems \cite{book}. Its
Hamiltonian is
\be
H=-t\sum_{j,\sigma}c^\dagger_{j,\sigma}c_{j+1,\sigma}+
c^\dagger_{j+1,\sigma}c_{j,\sigma}
+U\sum_j n_{j,\uparrow}\ n_{j,\downarrow}-\mu\sum_j n_j
-B\sum_j [n_{j,\uparrow}-n_{j,\downarrow}]\ ,
\label{HHubb}
\ee
where $n_{j,\sigma}=c^\dagger_{j,\sigma}c_{j,\sigma}$ and $n_j=
n_{j,\uparrow}+n_{j,\downarrow}$. In the following discussion the
magnetic field $B$ will be set to zero, but we will reinstate it in
the calculations in sections \ref{sec:hs}-\ref{sec:thres2}.
The Hubbard model is solvable by
Bethe Ansatz \cite{lw} and many exact results are available in the
literature \cite{book}. Of particular interest in view of experimental
applications are dynamical response functions such as the
single-particle spectral function
\bea
A(\omega,q)&=&-\frac{1}{\pi}\ {\rm Im}\ G_{\rm ret}(\omega,q),\nn
G_{\rm ret}(\omega,q)&=&-i\int_0^\infty dt\ e^{i\omega t}\sum_l e^{-iqla_0}
\langle 0|\{c_{j+l,\sigma}(t),\ c^\dagger_{j,\sigma}\}|0\rangle.
\label{specfun}
\eea
The spectral function is measured in angle-resolved photoemission
experiments. Such measurements on the quasi-1D organic conductor
TTF-TCNQ have been interprested in terms of $A(\omega,q)$ of the 1D
Hubbard model \cite{TTF,jeckel}. While high quality numerical results
are available from dynamical density matrix renormalization group
computations \cite{jeckel,jeckel2}, it is so far not possible to
calculate \fr{specfun} analytically from the exact solution. However,
using a field theory approach it is possible to determine low-energy
properties exactly. In particular, the singularity as a function of
$\omega$ at the Fermi wave number can be obtained using Luttinger
liquid theory \cite{Lutt}. The low-energy physics of the Hubbard model
in zero magnetic field is described by a spinful Luttinger liquid with
Hamiltonian $H=H_c+H_s$, where \cite{boso,book}
\bea
H=\sum_{\a=c,s}\frac{v_\a}{2\pi}\int dx\left[
\frac{1}{K_\a}\left(\frac{\partial \Phi_\a}{\partial x}\right)^2
+K_\a\left(\frac{\partial \Theta_\a}{\partial x}\right)^2\right]
+{\rm irrelevant\ operators}.
\label{HFT}
\eea
Here $K_s=1$ (we are concerned with the $B=0$ case for the time being) 
and the spin and charge velocities $v_{c,s}$ as well as the Luttinger
parameter $K_{c}$ are known functions of the density and interaction
strength \cite{book}. The Bose fields $\Phi_\a$ and the dual fields
$\Theta_\a$ fulfil the commutation relations
\be
\left[\Phi_\a(x),\frac{\partial\Theta_\b(y)}{\partial
    y}\right]=i\pi\delta_{\a\b}\delta(x-y).
\label{comm}
\ee
The spectrum of low-lying excitations (relative to the ground state)
in a large but finite system of size $L$ is given by
\cite{FSspectrum,book} 
\bea
\Delta E&=&\frac{2\pi  v_c}{L}
\left[\frac{(\Delta N_c)^2}{4\xi^2}
+\xi^2\Big(D_c+\frac{D_s}{2}\Big)^2+N_c^++N_c^-\right]
+\frac{2\pi  v_s}{L}
\left[\frac{\left(\Delta {N}_s-\frac{\Delta {N}_c}{2}\right)^2}{2}
+\frac{D_s^2}{2}+N_s^++N_s^-\right], \nn
\Delta P&=&\frac{2\pi }{L}
\left[\Delta N_c D_c+\Delta N_s D_s+N_c^+-N_c^-+N_s^+-N_s^-\right]
+2k_F(2D_c+D_s)\ ,
\label{ECFT}
\eea
where $\Delta N_\a$, $D_\a$ and $N_\a^\pm$ are integer or half-odd integer
``quantum numbers'' subject to the selection rules
\be
N_\a^\pm \in \mathbb{N}_0\ ,\quad \Delta N_\a\in \mathbb{Z}\ ,
\quad
D_c=\frac{\Delta N_c+\Delta N_s}{2}\text{mod}\ 1\ ,\quad
D_s=\frac{\Delta N_c}{2}\text{mod}\ 1.
\ee
We note that the form of the finite-size corrections implies that in
the finite volume the spin and charge sectors are not independent but
are in fact coupled through the boundary conditions of the fields
$\Phi_\a$, $\Theta_\a$.
Neglecting the effects of the
irrelevant operators in \fr{HFT} makes it possible to calculate
$A(\omega,q)$ at low energies \cite{Lutt,boso,book}. The spectral
function is found to exhibit singularities following the dispersions
of the collective spin (``spinon'') and charge (``holon'')
excitations. The exponents characterizing these singularities are
given in terms of the quantum numbers $\Delta N_\a$, $D_\a$ and $N_\a^\pm$.

\begin{figure}[ht]
\begin{center}
(a)
\epsfxsize=0.4\textwidth
\epsfbox{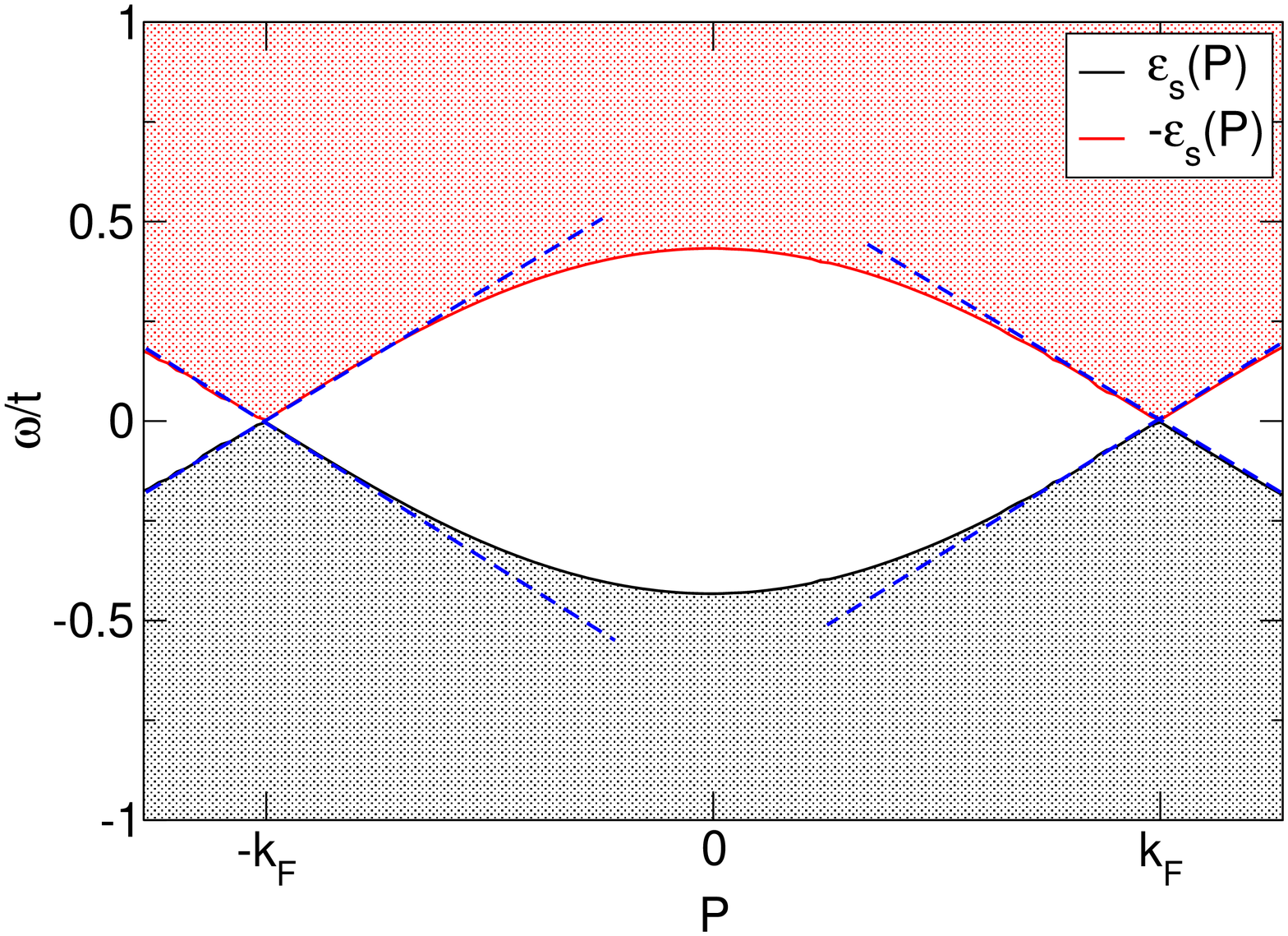}
\qquad(b)
\epsfxsize=0.4\textwidth
\epsfbox{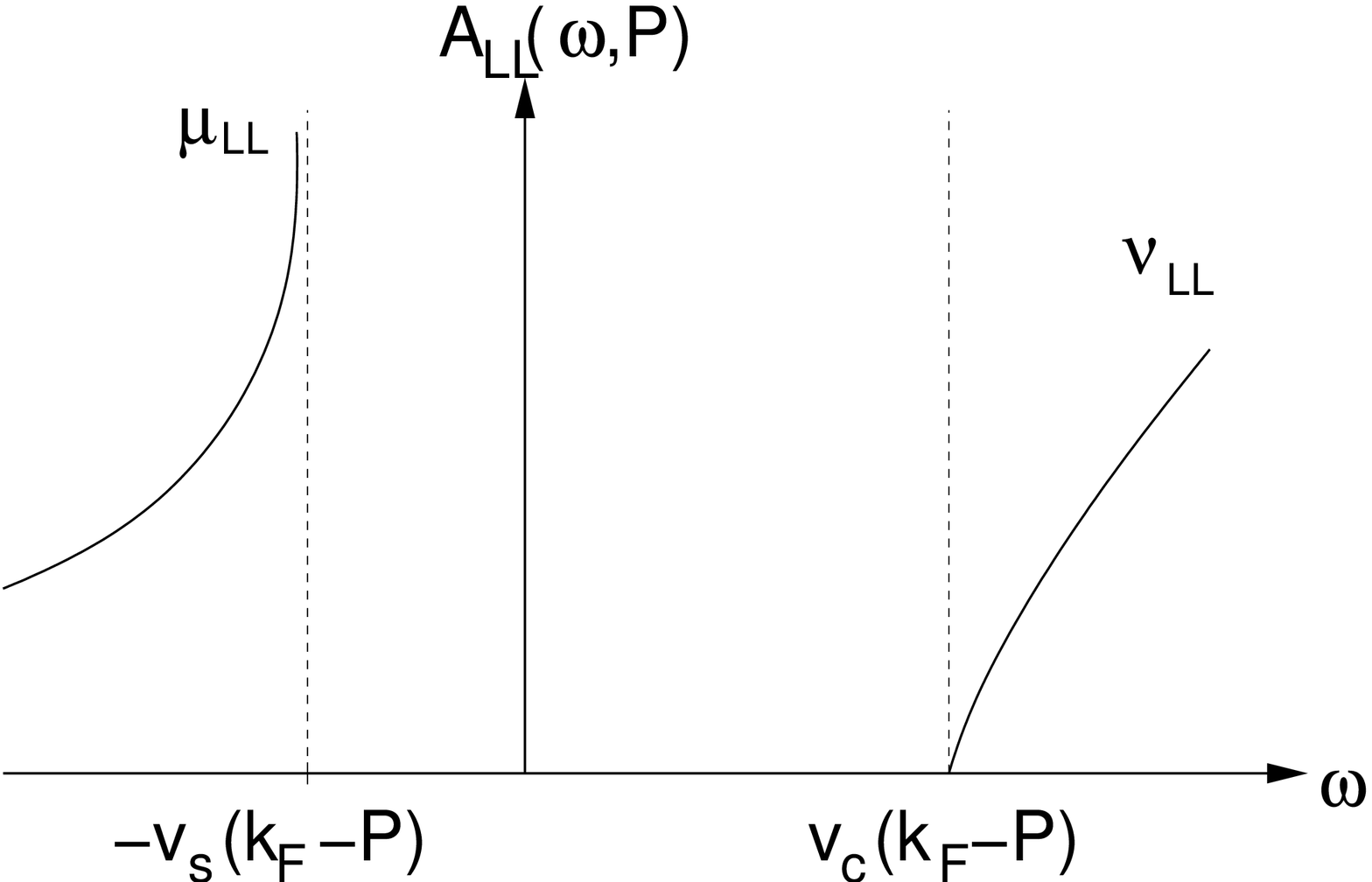}
\end{center}
\caption{(a) Support of the single-particle spectral
function. $A(\omega,P)$ is non-vanishing in the shaded
areas. $\eps_s(P)$ is the dispersion relation of the collective spin
excitations (``spinons''). The blue lines depict the linearized
dispersions $\pm v_s(P-k_F)$ and $\pm v_s(P+k_F)$ that underlie the
Luttinger liquid approximation.
(b) Structure of the single-particle spectral function in the
Luttinger liquid approximation for momenta close to $k_F$ and
$P<k_F$. There is a singularity at negative frequencies
$\omega=-v_s(k_F-P)$ and a power-law ``shoulder'' at positive
frequencies $\omega=v_c(k_F-P)$.}
\label{fig:aoqLL}
\end{figure}

\begin{figure}[ht]
\begin{center}
\epsfxsize=0.4\textwidth
\epsfbox{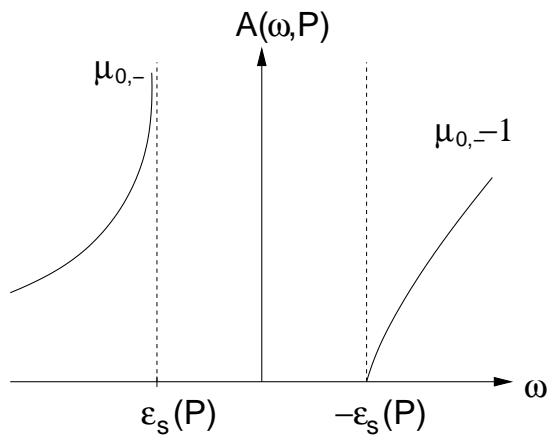}
\end{center}
\caption{Power-law singularity in the single-hole spectral function
  for $|P|<k_F$. There is a singularity at negative frequencies
  $\omega=\eps_s(P)$ with a momentum-dependent exponent $\mu_{0,-}$
  and a power-law ``shoulder'' at positive freuencies
  $\omega=-\eps_s(P)$. }
\label{fig:aoq}
\end{figure}

In a series of recent works \cite{work1,work2,work3,zgc1,work4,work5,work6,work7,zgc2,work8,Affleck2,zgc3} it was demonstrated for the
case of a spinless fermions, that neglecting the irrelevant operators
perturbing the Luttinger liquid Hamiltonian leads in general to
incorrect results for singularities in response functions. Using a
mapping to a Luttinger liquid coupled to a mobile impurity and taking
the leading irrelevant operators into account non-perturbatively it is
possible to determine the exact singularities in response
functions \cite{work1,work2,work3,zgc1,work4,work5,work6,work7,zgc2,work8,Affleck2,zgc3,austen}. 
Crucially, these singularities are generally
\emph{momentum dependent}. Two recent preprints have addressed the
generalization to spinful fermions \cite{Pereira,SIG}. In particular,
Ref.~\onlinecite{SIG} derives expressions for the exponents
$\mu_{n,\pm}$ characterizing the the singularities of the
single-particle spectral function \fr{specfun}. The resulting spectral
function is depicted in Fig. \ref{fig:aoq}.

Our goal is to provide an exact expression for the threshold exponents
$\mu_{0,-}$ in terms of the microscopic parameters entering the
Hamiltonian, i.e. $U$, $\mu$ and $B$ in the case of the Hubbard
model. This is achieved by determining the finite-size corrections to
low-lying  energy levels {\sl in presence of a high-energy
  excitation}. Comparing the results obtained from the Bethe Ansatz
solution of the Hubbard model as well as the closely related
Yang-Gaudin model to field theory predictions, we are able to 
derive explicit expressions for the threshold exponents. In the
Yang-Gaudin case our explicit results agree with the relations of
spectrum and exponents proposed in Ref.\cite{SIG} for
Galilei-invariant models\endnote{We note that these relations by
themselves do not provide explicit expressions of the threshold
exponents in terms of the microscopic parameters entering the
Hamiltonian.}. 

There is one caveat for the case of the Hubbard model. As for any
lattice model it is possible to generate low-lying excitations
for any momentum by combining multiple Umklapp processes if the band
filling is incommensurate. A completely analogous situation is
encountered for spinless fermions \cite{Affleck2}. Hence, for
incommensurate fillings no thresholds exist. While we would still
expect the spectral function to feature peaks asociated with the
thresholds of particular excitations, in particular those involving
small numbers of holons and spinons, these peaks will no longer
correspond to singularities. In order to circumvent this problem we
will consider only the case of commensurate fillings in the Hubbard
model. 

The outline of this paper is as follows. In section \ref{sec:BA} we
briefly review the Bethe Ansatz description of the ground state of the
Hubbard model. In section \ref{sec:exc} we present the excitations
that give rise to low-energy thresholds around the Fermi momentum in
the Hubbard model. In sections \ref{sec:thres1} and \ref{sec:thres2} 
we determine the finite-size spectrum of excited states describing
these thresholds. In section \ref{sec:FT} we relate these results to
the field theory treatment of the threshold problem and extract
threshold exponents. In section \ref{sec:egas} we summarize the
analogous results for the Yang-Gaudin model.

\section{Bethe Ansatz equations for the Hubbard Model}
\label{sec:BA}
The logarithmic form of the Bethe ansatz equations of the Hubbard
model for $N$ electrons out of which $M$ have spin down (for real
solutions only) is 
\begin{eqnarray} 
&&     k_j L  =  2 \pi I_j - \sum_{\alpha = 1}^{M}
                 \theta \left(
		 \frac{\sin k_j - \Lambda_\alpha}{u} \right),\quad
               j=1,\ldots,N\ ,
		 \nn
&&     \sum_{j=1}^{N} \theta \left(
		 \frac{\Lambda_\alpha - \sin k_j}{u} \right)  = 
		 2 \pi J_\alpha+
		 \sum_{\beta = 1}^{M}
		 \theta \left(
		 \frac{\Lambda_\alpha - \Lambda_\beta}{2u} \right)\ ,\
\alpha=1,\ldots,M.
\label{BAE}
\end{eqnarray}
Here 
\be
u=\frac{U}{4t},
\ee
the length of the lattice $L$ is taken to be even, $\theta(x)
=2 \arctan(x)$ and $I_j$, ${J}_\alpha$ are integer or half-odd integer
numbers that arise due to the multivaluedness of the logarithm. They
are subject to the ``selection rules''
\be
I_j\ {\rm is}\ \bigg\{\begin{array}{l l}
{\rm integer} &\text{if}\ M\ \text{is even}\\
{\rm half-odd\ integer} &\text{if}\ M\ \text{is odd},\\
\end{array}
\label{int-hoi1}
\end{equation}
\be
J_\a\ {\rm is}\ \bigg\{\begin{array}{l l}
{\rm integer} &\text{if}\ N-M\ \text{is odd}\\
{\rm half-odd\ integer} &\text{if}\ N-M\ \text{is even},\\
\end{array}
\label{int-hoi2}
\end{equation}
\be
-\frac{L}{2}<I_j\leq\frac{L}{2}\ ,\qquad
|J_\a|\leq \frac{1}{2}(N-M-1)\ .
\label{ranges}
\ee
The energy (in units of $t$) and momentum of such Bethe ansatz states are
\be
E=uL+2BM-\sum_{j=1}^N\left[2\cos(k_j)+\mu+2u+B\right] ,\qquad
P=\sum_{j=1}^N k_j\equiv\frac{2\pi}{L}\left[
\sum_{j=1}^NI_j+\sum_{\alpha=1}^MJ_\alpha\right].
\label{EP}
\ee
\subsection{Ground State below half-filling}
Following section 7.7 of [\onlinecite{book}] we now take
consider the ground state below half filling.
We have 
\be
N=N_{\rm GS}\ ,\quad 
M=M_{\rm GS},
\ee
where we take $N_{\rm GS}=2\times {\rm odd}$ and $M_{\rm GS}={\rm
  odd}$. We note that in zero magnetic field we have $M_{\rm
  GS}=\frac{N_{\rm GS}}{2}$. Our choice for $N$ and $M$ implies that
$I_j$ are half-odd integers and $J_\alpha$ are integers. In the ground
state all vacancies are filled symmetrically around zero
\bea
I_j&=&j-\frac{N_{\rm GS}+1}{2}\ ,\qquad j=1,\ldots,N_{\rm GS}\ ,\nn
J_\alpha&=&\alpha-\frac{M_{\rm GS}+1}{2}\ ,\qquad
\alpha=1,\ldots,M_{\rm GS}\ . 
\label{integersGS}
\eea
It follows from \fr{EP} that the ground state momentum is zero. The
bulk ground state energy can be expressed in terms of the solution of
the following set of coupled integral equations
\begin{eqnarray}
\rho_c(k)&=&\frac{1}{2\pi}+\cos k \int_{-A}^A 
d\l\ \fa{1}{\sin k-\l}
\ \rho_s(\Lambda)\ ,
\label{densGS1}\\
\rho_s(\Lambda)&=&\int_{-Q}^Q
dk\ \fa{1}{\l-\sin k}
\ \rho_c(k)-\int_{-A}^A
d\lp\ \fa{2}{\l-\lp}
\rho_s(\lp)\ ,
\label{densGS2}
\end{eqnarray}
where
\be
a_n(x)=\frac{1}{2\pi}\frac{2nu}{(un)^2+x^2}.
\label{an}
\ee
The integrated densities yield the total number of electrons per site
and the number of electrons with spin down per site respectively
\begin{equation}
\int_{-Q}^Q dk \rho_c(k)=\frac{N_{\rm GS}}{L}\equiv n_c\ ,\qquad
\int_{-A}^A d\Lambda \rho_s(\Lambda)=\frac{M_{\rm
    GS}}{L}=\frac{N_\da}{L}\equiv n_s\ . 
\label{DM}
\end{equation}
The integration boundaries $Q$ and $A$ can be fixed in terms of
$N_{\rm GS}$ and $M_{\rm GS}$ by these equations. Alternatively one
can define {\sl dressed   energies} by
\begin{eqnarray}
\eps_c(k)&=&-2\ \cos k-\mu-2u-B+\int_{-A}^A
d\l\ \fa{1}{\sin k-\l}
\ \eps_s(\Lambda)\ ,\fnn
\eps_s(\Lambda)&=&2B+\int_{-Q}^Q
dk\ \cos(k)\ \fa{1}{\sin k-\l}
\eps_c(k)
-\int_{-A}^A
d\lp\ \fa{2}{\l-\lp}
\ \eps_s(\lp)\ .
\label{kappaeps}
\end{eqnarray}
Here the integration boundaries $\pm Q$ and $\pm A$ are by definition
the points at which the dressed energies switch sign, so that they
are determined as functions of the chemical potential and the magnetic
fields via the conditions
\begin{equation}
\eps_c(\pm Q)=0\ ,\qquad
\eps_s(\pm A)=0\ .
\label{QA}
\end{equation}
The bulk ground state energy per site is 
\begin{eqnarray}
e_{\rm GS}&=&\int_{-Q}^Q dk\ (-2\cos k-\mu-B -2u)\ \rho_c(k)
+2Bn_s+u
=\int_{-Q}^Q \frac{dk}{2\pi}\ \eps_c(k)\ +u.
\label{gsE}
\end{eqnarray}

\section{Excitations with Charge $-e$ and Spin $\frac{1}{2}$} 
\label{sec:exc}
We now consider an excitation over the ground state with the quantum
numbers of a hole with spin down. As the total charge must be one less
than in the ground state we must have
\be
N=N_{\rm GS}-1.
\label{dq}
\ee
Recalling that the $z$-component of total spin quantum number is
$S^z=\frac{N-2M}{2}$ we see that relative to the ground state we must have
\be
\delta S^z=\frac{2l+1}{2}\ ,\quad l\in\mathbb{N}_0.
\label{dsz}
\ee
Equation \fr{dsz} requires an explanation. As shown in
\cite{hwt} the Bethe ansatz states only provide \emph{highest-weight
  states} of the SO(4) symmetry \cite{so4} of the Hubbard model. The
corresponding SO(4) multiplet is then obtained by acting with lowering
operators. As a result, for any excitation given by the Bethe ansatz with
$\delta S^z=\frac{2l+1}{2}$ for non-negative integers $l$ we can
construct an excited state with $\delta S^z=\frac{1}{2}$ by acting
with the spin lowering operator. In presence of a magnetic field this
shifts the energy by $-2Bl$.  

\subsection{Holon-Spinon Excitation} 
\label{sec:hs}
\begin{figure}[ht]
\begin{center}
\epsfxsize=0.45\textwidth
\epsfbox{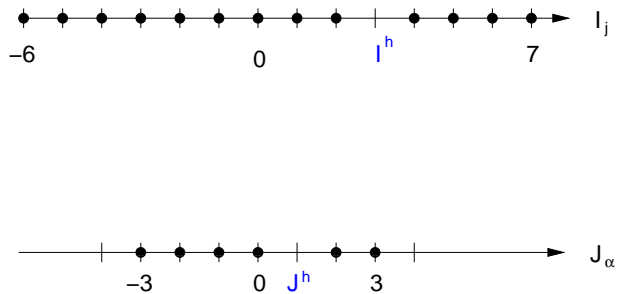}
\end{center}
\caption{Distribution of integers for the holon-spinon excitation
for $N_{\rm GS}=14$, $M_{\rm GS}=7$. The positions of the two holes
are $I^h$ and $\Lambda^h$ respectively. We have shown the
configuration where the $I_j$' range from 
$\frac{N_{\rm  GS}}{2}-1$ to $\frac{N_{\rm GS}}{2}$, but we equally
well can choose them 
$\frac{N_{\rm  GS}}{2}\leq I_j\leq \frac{N_{\rm GS}}{2}-1$. The two
possibilities give rise to the two signs in \fr{EPhs}.
} 
\label{fig:integers}
\end{figure}

\begin{figure}[ht]
\begin{center}
(a)\epsfxsize=0.46\textwidth
\epsfbox{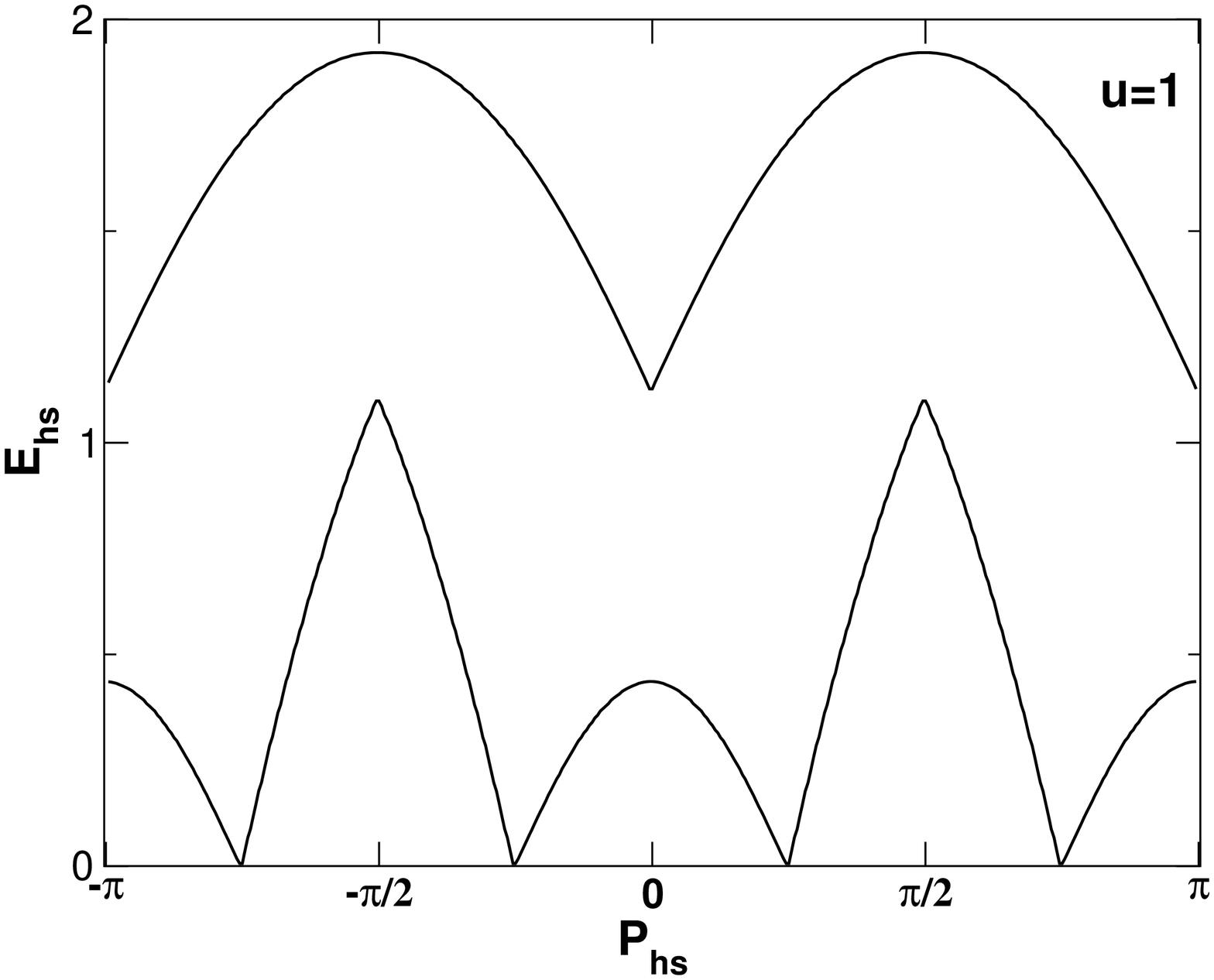}\quad
(b)\epsfxsize=0.46\textwidth
\epsfbox{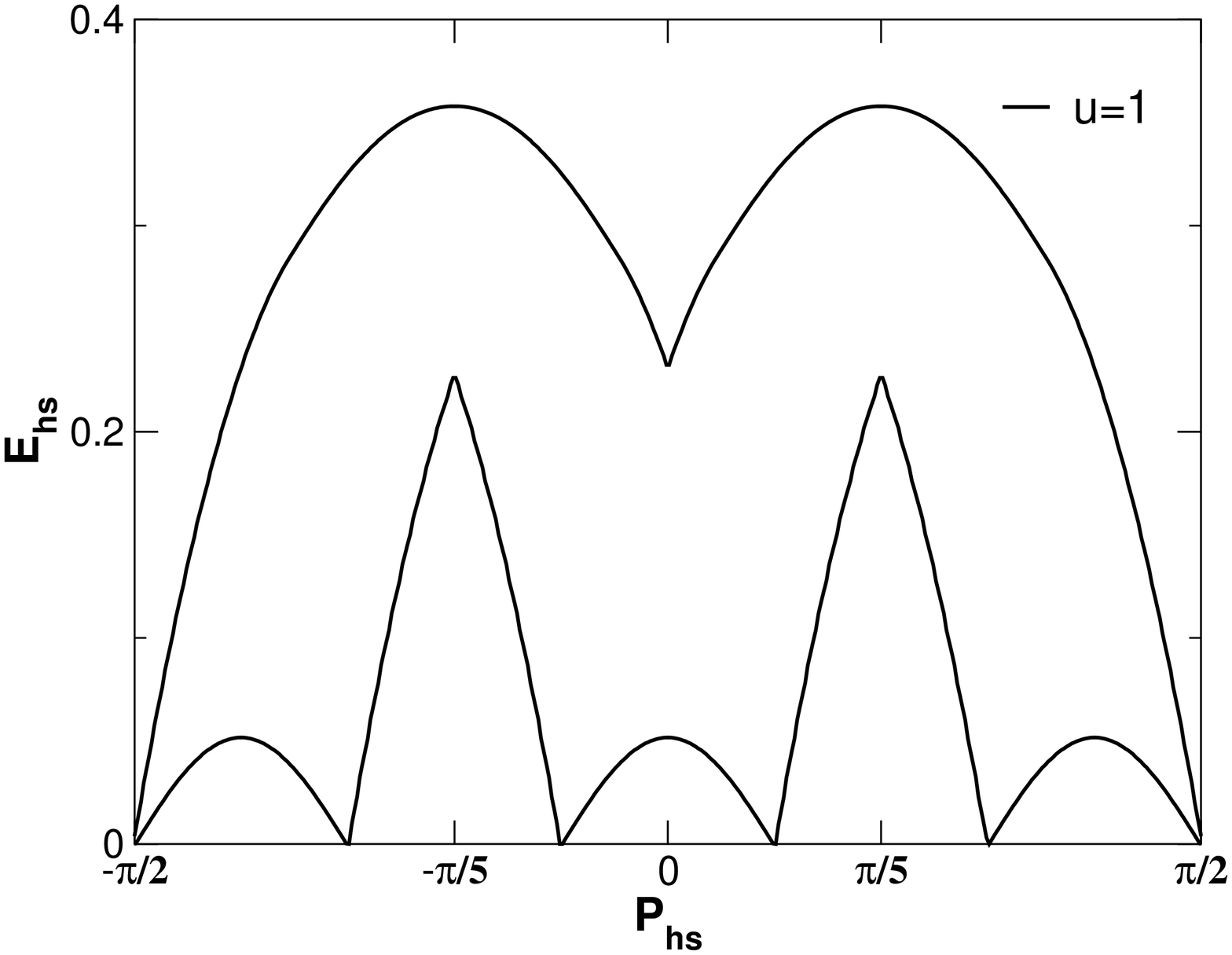}
\end{center}
\caption{Boundaries of the holon-spinon excitation continuum in zero
  magnetic field for (a) $u=1$ and density $n_c=0.5$ and (b) for $u=1$
  and $n_c=0.2$.}  
\label{fig:hs}
\end{figure}

The simplest excitation with the quantum numbers \fr{dsz}, \fr{dq}
is {\sl two-parametric} \cite{book} and obtained by setting
\be
N=N_{\rm GS}-1\ ,\quad M=M_{\rm GS}-1,
\ee
in the Bethe ansatz equations \fr{BAE}. It then follows from
\fr{int-hoi1}, \fr{int-hoi2} that both $I_j$ and $J_\a$ are
integers. In order to see that we are dealing with a two-parametric
excitation we consider the number of vacancies for the integers $I_j$
and $J_\a$ \fr{ranges}. As 
\be
-\frac{L}{2}< I_j\leq\frac{L}{2},
\ee
there is the same number of vacancies for the $I_j$'s as in the
ground state, but we have one fewer integer, i.e. one additional
``hole''. Similarly we have 
\be
|J_\a|\leq\frac{1}{2}(N_{\rm GS}-M_{\rm GS}-1),
\ee
which tells us that we have the same number of vacancies as in the
ground state but one root less, which leaves one ``hole''. It
is useful to plot the distribution of integers. This is done in
Fig.\ref{fig:integers}. 
The energy and momentum of the ``holon-spinon'' excitation described
above are determined in Chapter 7 of Ref.\onlinecite{book}
\bea
E_{hs}&=&-\eps_c(k^h)-\eps_s(\Lambda^h)\ ,\nn
P_{hs}&=&-p_c(k^h)-p_s(\Lambda^h)\pm \pi n_c,
\label{EPhs}
\eea
where the dressed energies are defined in \fr{kappaeps} and the
dressed momenta are
\bea
p_c(k)&=&2\pi\int_0^k dk'\ \rho_c(k')\ ,\nn
p_s(\l)&=&\pi(n_c-n_s)-2\pi\int_\l^\infty d\l\ \rho_s(\l).
\label{dressedmta}
\eea
The extra contribution $\pm \pi n_c$ in \fr{EPhs} arises as the
distribution of $I_j$'s is asymmetric with respect to the origin.
In zero magnetic field the equation for $p_s(\l)$ can be simplified
\be
p_s(\Lambda)\Bigl|_{B=0}=\frac{\pi n_c}{2}-2\int_{-Q}^Q dk\ {\rm arctan}
\left[\exp\left(-\frac{\pi}{2u}(\Lambda-\sin(k)\right)\right]\rho_c(k).
\ee
In Fig.\ref{fig:hs} we plot the upper and lower boundaries of the
holon-spinon continuum for two different band fillings in zero
magnetic field. We see that as expected there are soft modes around
$\pm k_F=\frac{\pi n_c}{2}$.  
\subsection{Holon - 3 Spinon Excitation} 
\label{sec:h3s}
\begin{figure}[ht]
\begin{center}
(a)\epsfxsize=0.46\textwidth
\epsfbox{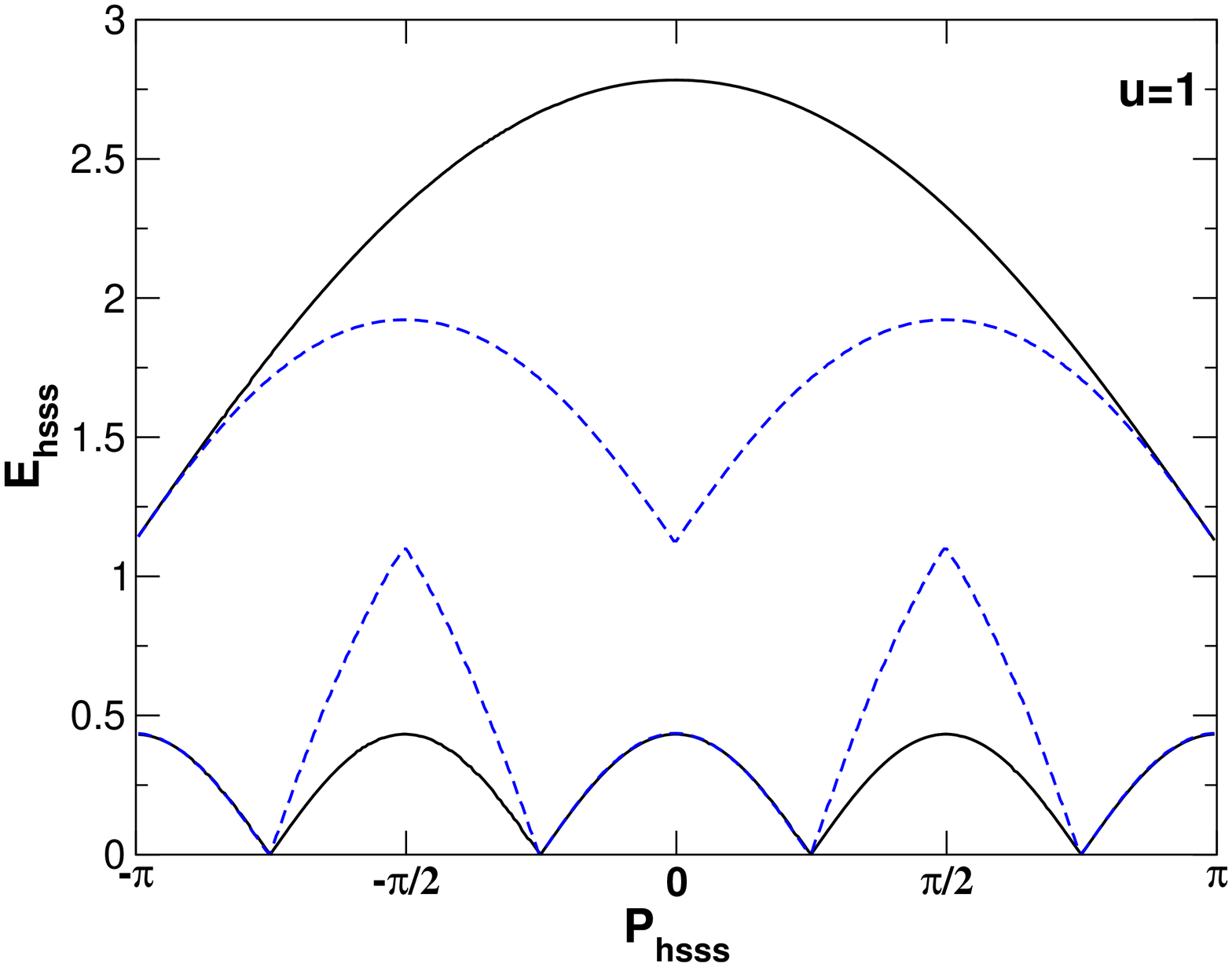}\quad
(b)\epsfxsize=0.46\textwidth
\epsfbox{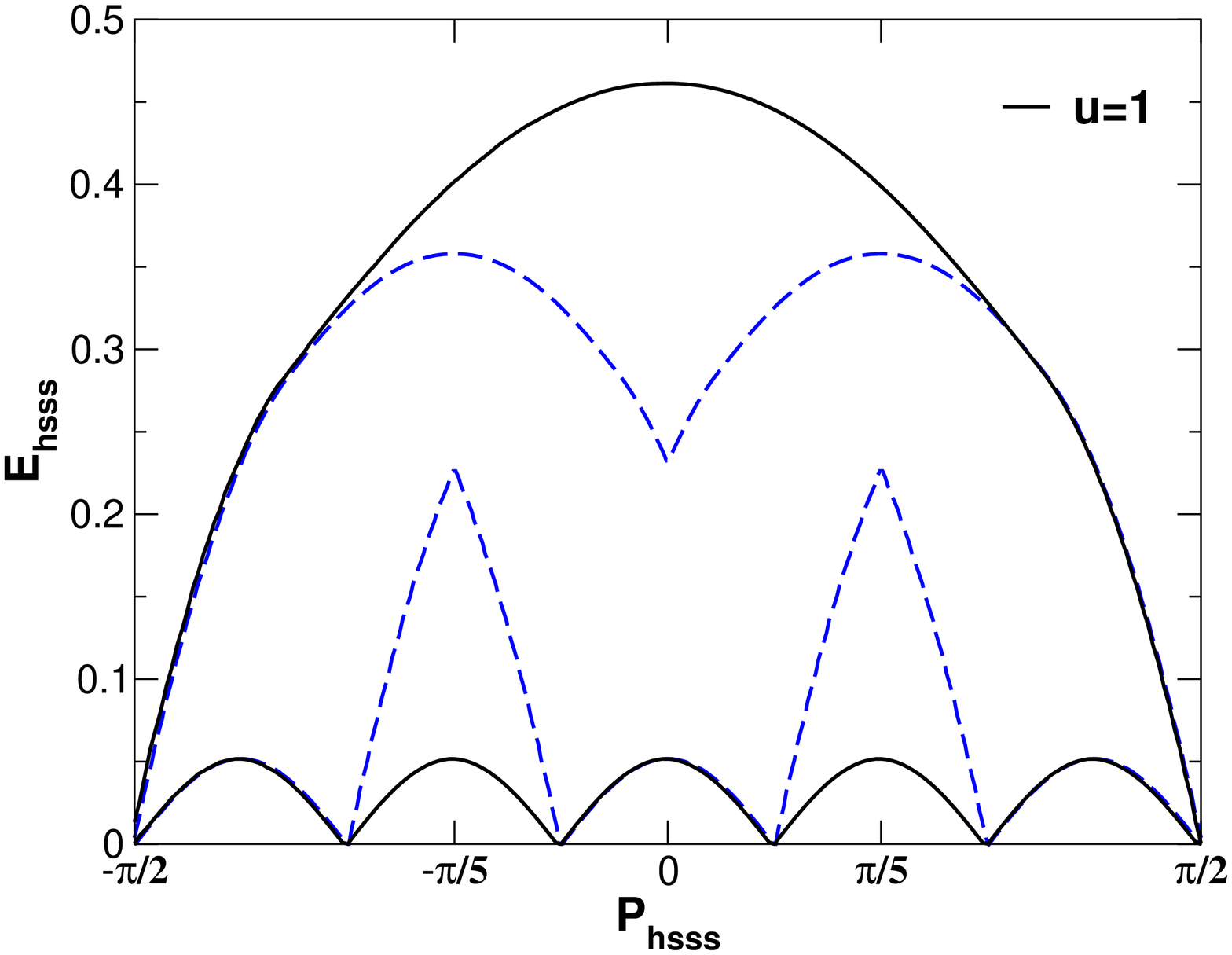}
\end{center}
\caption{Boundaries of the holon - 3 spinon excitation continuum in
  zero magnetic field for (a) $u=1$ and density $n_c=0.5$ and (b) for
  $u=1$ and $n_c=0.2$. The boundaries of the holon-spinon continua 
are depicted by the dashed blue lines.} 
\label{fig:hsss}
\end{figure}

Other kinds of excitations with the quantum numbers \fr{dsz}, \fr{dq}
involve more than two ``elementary'' excitations. As we are interested
in the threshold of the spectral function, we focus on excitations
that lead to the smallest possible energy for a given momentum. This
leads us to consider holon - 3 spinon excitations characterized by
a solution of the Bethe ansatz equations \fr{BAE} with
\be
N=N_{\rm GS}-1\ ,\quad M=M_{\rm GS}-2.
\ee
It follows from \fr{int-hoi1}, \fr{int-hoi2} that both $I_j$
and $J_\a$ are half-odd integers. In order to see that we are dealing with a
four-parametric excitation we consider the number of vacancies
for the integers $I_j$ and $J_\a$ \fr{ranges}. As
\be
-\frac{L}{2}< I_j\leq\frac{L}{2},
\ee
there is the same number of vacancies for the $I_j$'s as in the
ground state, but we have one fewer integer, i.e. one additional
``hole''. Similarly we have 
\be
|J_\a|\leq\frac{1}{2}(N_{\rm GS}-M_{\rm GS}),
\ee
which tells us that we have one more vacancy than in the
ground state but two roots less, which amounts to three ``holes''.
The energy and momentum of this excitation are
\bea
E_{hsss}&=&-\eps_c(k^h)-\sum_{j=1}^3
\eps_s(\Lambda^h_j)\ ,\nn
P_{hsss}&=&-p_c(k^h)-\left(\sum_{j=1}^3 p_s(\Lambda^h_j)\right),
\label{EPhsss}
\eea
where the dressed energies and momenta are defined in \fr{kappaeps}
and \fr{dressedmta}, respectively. We note that there is no aditional
constant contribution to the momentum because both the $I_j$ and
$J_\a$ are half odd integers, leading to symmetric distributions of
roots in the absence of holes.
In Fig.\ref{fig:hsss} we plot the upper and lower boundaries of the
holon - 3 spinon continuum for two different band fillings in zero
magnetic field. We see that in part of the Brillouin zone the holon -
3 spinon threshold is lower in energy than the holon-spinon
continuum. 

\subsection{Thresholds in Zero Magnetic Field}
We are now in a position to determine the lower boundaries of the
holon-spinon and holon - 3 spinon continua in the vicinity of
$k_F$. We find that the lower boundary is obtained as follows:
\begin{itemize}
\item{} \underline{$0<P_{hs}<k_F$}

Here the threshold is identical for both types of excitation
considered above. The leading singularity of the spectral function is
given by contribution of the holon-spinon continuum. The threshold is
obtained as follows: we choose the plus sign in \fr{EPhs} and place
the holon at its right ``Fermi point'', i.e. 
\be
I^h=\frac{N_{\rm GS}}{2}\ ,\qquad k^h=Q.
\label{ihole}
\ee
It therefore carries zero energy and momentum $-\pi n_c$. For
$P_{hs}=k_F$ the spinon has rapidity $\Lambda^h=-\infty$, which
corresponds to momentum $k_F=\frac{\pi n_c}{2}$. Taking the spinon
rapidity from $\Lambda^h=-\infty$ to $\Lambda^h=0$ then traces out the
lower boundary of the holon-spinon continuum for $0<P_{hs}<k_F$.

\item{} \underline{$2k_F>P_{hs}>k_F$}

Here the threshold is given by the holon - 3 spinon continuum. It is
obtained by placing the holon at its left ``Fermi point''
\be
I^h=-\frac{N_{\rm GS}-1}{2}\ ,\qquad k^h=-Q.
\label{ihole2}
\ee
It therefore carries zero energy and momentum $\pi n_c=2k_F$. Two of the
spinons are placed at their left and right ``Fermi points''
respectively
\be
J^h_{1,2}=\pm\frac{M_{\rm GS}}{2}\ ,\qquad \Lambda^h_{1,2}=\pm \infty.
\label{Jholes}
\ee
Their contributions to the total momentum cancel. To obtain total
momentum $P_{hsss}=k_F$ the third spinon is taken to have rapidity
$\Lambda^h=\infty$, which corresponds to momentum $-k_F=-\frac{\pi
  n_c}{2}$. Taking the spinon rapidity from $\Lambda^h=\infty$ to
$\Lambda^h=0$ then traces out the lower boundary of the holon - 3 spinon
continuum for $k_F<P_{hsss}<2k_F$. 
\end{itemize}

At this point an obvious question to ask is whether other excitations
may lead to even lower thresholds. As we have restricted ourselves to
commensurate band fillings and zero magnetic field the answer close to
$k_F$ is negative. This follows from the general expression for the
low-energy part of the spectrum \fr{ECFT}. This strongly suggests that
the threshold of the spectral function is given by the two excitations
given above and we will assume this to be the case in what follows.

\subsection{Holon-Spinon Threshold  in Zero Magnetic Field}
It is conceivable that the spectral function $A(\omega,q)$ could
exhibit features associated with the theshold of the holon-spinon
excitation in the momentum range where the threshold of $A(\omega,q)$
occurs at a lower frequency. We therefore determine the threshold of
the holon-spinon excitation in zero field in the 
momentum range $2k_F>P_{hs}>k_F$. It is obtained by choosing the
plus sign in \fr{EPhs} and place the spinon at its ``Fermi point'', i.e. 
\be
J^h=-\frac{N_{\rm GS}-2}{4}\ ,\qquad \Lambda^h=-\infty.
\ee
It therefore carries zero energy and momentum 
$\frac{\pi n_c}{2}$. For
$P_{hs}=k_F$ the holon has rapidity $k^h=Q$, which
corresponds to momentum $-\pi n_c$. Reducing the holon
rapidity from $k^h=Q$ traces out the lower boundary of the
holon-spinon continuum for $k_F<P_{hs}<2k_F$. 
\section{Threshold at $0<P<k_F$ for the Holon-Spinon Excitation}
\label{sec:thres1}
In the following we consider the threshold at $0<P<k_F$ for the
holon-spinon excitation. By virtue
of \fr{ihole} we are dealing with a distribution of integers $I_j$
that is symmetric around zero. The free parameter is then the spinon 
rapidity $\Lambda^h$. More precisely, the distributions of integers
are
\bea
I_j&=&-\frac{N_{\rm GS}}{2}+j\ ,\quad 1\leq j\leq N_{\rm GS}-1\ ,\nn
J_\a&=&\begin{cases}
-\frac{M_{\rm GS}+1}{2}+\a & \text{if}\ 1\leq \a<\frac{M_{\rm GS}+1}{2}+J^h\cr
-\frac{M_{\rm GS}+1}{2}+\a+1 & \text{if}\ \frac{M_{\rm GS}+1}{2}+J^h\leq
\a\leq M_{\rm GS}-1\cr
\end{cases}.
\eea
Our goal is to determine the finite-size corrections to the energy of
this excitation in the limit 
\be
L,J^h,N_{\rm GS}, M_{\rm GS}\to \infty\ ,\qquad
\frac{J^h}{L},\frac{N_{\rm GS}}{L},\frac{M_{\rm GS}}{L}\ \text{fixed}.
\ee

\subsection{Finite-Size Corrections}
Our starting point are the Bethe ansatz equations \fr{BAE} for the
holon-spinon excitation where the holon sits at the Fermi momentum of
the $k_j$'s. It is convenient to write them in terms of \emph{counting
functions}
\begin{eqnarray}
z_c(k)  &=&  k+ \frac{1}{L} \sum_{\alpha = 1}^{M}
                 \theta \left(
		 \frac{\sin k - \Lambda_\alpha}{u} \right),\\
z_s(\l)&=&\frac{1}{L}\sum_{j=1}^{N} \theta \left(
		 \frac{\Lambda - \sin k_j}{nu} \right)  -
		\frac{1}{L} \sum_{\beta = 1}^{M}
		 \theta \left(
		 \frac{\Lambda- \Lambda_\beta}{2u} \right)\ .
\label{CF}
\end{eqnarray}
The Bethe Ansatz equations then read
\be
z_c(k_j)=\frac{2\pi I_j}{L}\ ,\qquad
z_s(\l_\a)=\frac{2\pi J_\a}{L}\ .
\ee

We now turn these into integral equations by means of the
Euler-Maclaurin sum formula
\be
\frac{1}{L}\sum_{n=n_1}^{n_2}f\Big(\frac{n}{L}\Big)=
\int_{\frac{n_-}{L}}^{\frac{n_+}{L}}dx\
f(x)+\frac{1}{24L^2} \left[f'\Big(\frac{n_-}{L}\Big)
-f'\Big(\frac{n_+}{L}\Big)\right]+\ldots,
\label{EMcL}
\ee
where
\be
n_+=n_{2}+\frac{1}{2}\ ,\quad
n_-=n_{1}-\frac{1}{2}.
\ee
For the specific exitation we are considering we have
\bea
I_\pm=\pm\frac{N_{\rm GS}-1}{2}\ ,\qquad
J_\pm=\pm\frac{M_{\rm GS}}{2}.
\eea
Using \fr{EMcL} in \fr{CF} and then changing variables
from $x$ (i.e. integers divided by $L$) to the rapidities we arrive at 
\bea
z_c(k)&=&k+\int_{A_-}^{A_+}d\l\ \rho_s(\l)\ 
\th\Bigl(\frac{\sin k-\l}{u}\Bigr)-\frac{1}{L}
\th\Bigl(\frac{\sin k-\l^h_L}{u}\Bigr)\nn
&+&\frac{1}{24L^2}\left[
\frac{a_1(\sin k-A_+)}{\rho_s(A_+)}
-\frac{a_1(\sin k-A_-)}{\rho_s(A_-)}\right]+o(L^{-2}),\\
z_s(\l)&=&\int_{Q_-}^{Q_+}dk\
\th\Bigl(\frac{\l-\sin k}{u}\Bigr)\rho_c(k)
-\int_{A_-}^{A_+}d\l'\ \rho_s(\l')\ 
\th\Bigl(\frac{\l-\l'}{2u}\Bigr)+\frac{1}{L}
\th\Bigl(\frac{\l-\l^h_L}{2u}\Bigr)\nn
&+&\frac{1}{24L^2}\left[
\frac{a_1(\l-\sin Q_+)\cos Q_+}{\rho_c(Q_+)}
-\frac{a_1(\l-\sin Q_-)\cos Q_-}{\rho_c(Q_-)}
-\frac{a_2(\l-A_+)}{\rho_s(A_+)}
+\frac{a_2(\l-A_-)}{\rho_s(A_-)}
\right]+o(L^{-2}).
\label{count}
\eea
Here $a_n(x)$ is given in \fr{an}, the integration boundaries are
fixed by 
\be
z_c(Q_\pm)=\frac{2\pi I_{\pm}}{L}\ ,\qquad
z_s(A_\pm)=\frac{2\pi J_{\pm}}{L},
\ee
and the root densities are defined by
\be
2\pi\rho_s(\l)=\frac{dz_s(\l)}{d\l}\ ,\qquad
2\pi\rho_c(k)=\frac{dz_c(k)}{dk}.
\label{bcs}
\ee
In addition there is the equation fixing the position of the hole
\be
z_s(\l^h_L)=\frac{2\pi J^h}{L}.
\ee
Here our notation makes the $L$-dependence of the rapidity of the hole
explicit. 
Taking derivatives of \fr{count} we obtain coupled linear integral
equations for the root densities $\rho_{c,s}$
\bea
\rho_c(k)&=&\frac{1}{2\pi}+\cos(k)\int_{A_-}^{A_+}d\l\ \rho_s(\l)\ 
a_1(\sin k-\l)-\frac{a_1(\sin k-\l^h_L)\cos k}{L}\nn
&&+\frac{\cos k}{24L^2}\left[
\frac{a_1'(\sin k-A_+)}{\rho_s(A_+)}
-\frac{a_1'(\sin k-A_-)}{\rho_s(A_-)}\right],\\
\rho_s(\l)&=&\int_{Q_-}^{Q_+}
a_1(\l-\sin k)\rho_c(k)
-\int_{A_-}^{A_+}d\l'\ \rho_s(\l')\ 
a_2(\l-\l')+\frac{1}{L}a_2(\l-\l^h_L)\nn
&&+\frac{1}{24L^2}\left[
\frac{a_1'(\l-\sin Q_+)\cos Q_+}{\rho_c(Q_+)}
-\frac{a_1'(\l-\sin Q_-)\cos Q_-}{\rho_c(Q_-)}
-\frac{a_2'(\l- A_+)}{\rho_s(A_+)}
+\frac{a_2'(\l- A_-)}{\rho_s(A_-)}
\right].
\label{dens}
\eea
In shorthand notations these can be written as
\be
\rho_\a=\rho_\a^{(0)}+\widehat{K}_{\a\beta}*\rho_\beta\ ,
\ee
where $*$ denotes convolution on the interval $[Q_-,Q_+]$
($[A_-,A_+]$) if $\beta=c$ ($\beta=s$). The components of the matrix
kernel are
\bea
K_{cc}(k,k')&=&0\ ,\qquad\qquad\quad\qquad K_{cs}(k,\l)=\cos(k)\ a_1(\sin k-\l)\ ,\nn
K_{sc}(\Lambda,k)&=&a_1(\sin k-\l)\ ,\qquad
K_{ss}(\l,\l')=-a_2(\l-\l').
\label{kernel}
\eea
It is useful to introduce a
unified notation for the integration boundaries
\be
X^\a_\pm=
\begin{cases}
Q_\pm & \text{if}\ \a=c\cr
A_\pm & \text{if}\ \a=s.
\end{cases}
\ee
Using the Euler-MacLaurin sum formula on the expression for the energy
\fr{EP} we obtain
\bea
E=Lu+L\sum_\a\int_{X^\a_-}^{X^\a_+} dz\ \eps_\a^{(0)}(z)\ \rho_a(z)
-\eps^{(0)}_s(\l^h_L)-\frac{1}{24L}\sum_\a
\frac{\eps^{(0)'}_\a(X^\a_+)}{\rho_\a(X^\a_+)}-
\frac{\eps^{(0)'}_\a(X^\a_-)}{\rho_\a(X^\a_-)},
\eea
where the prime denotes the derivative with respect to the argument
and $\eps^{(0)}_\a$ are the {\sl bare energies}, i.e. the driving
terms in \fr{kappaeps}
\be
\eps^{(0)}_c(k)=-2\cos(k)-\mu-2u-B\ ,\qquad
\eps^{(0)}_s(\l)=2B.
\label{bare}
\ee

Now we have to solve the system of equations \fr{dens} with boundary
conditions \fr{bcs} in by expanding both the densities and the
integration boundaries in inverse powers of $L$. As the integral
equations are linear we may proceed by first keeping the integration
boundaries general and only expanding the densities as
\bea
\rho_\a(z)=\rho_{\a,0}(z)+\frac{1}{L}\rho_{\a,1}(z)+\frac{1}{24L^2}
\sum_{\b,\sigma=\pm}\frac{f^{(\sigma)}_{\a\b}(z)}{\rho_\b(X^\b_\sigma)}.
\eea
The various parts then fulfil the integral equations
\bea
\rho_{\a,a}&=&\rho_{\a,a}^{(0)}+\widehat{K}_{\a\beta}*\rho_{\beta,a}\ ,
\quad a=0,1\ ,\label{inteq}\\
f^{(\sigma)}_{\a\b}&=&d^{(\sigma)}_{\a\b}+\widehat{K}_{\a\g}*f^{(\sigma)}_{\g\b},
\eea
where the driving terms are given by
\bea
\rho_{\a,0}^{(0)}(z)&=&\frac{\delta_{\a c}}{2\pi}\ ,\\
\rho_{\a,1}^{(0)}(z)&=&-K_{\a s}(z,\l^h_L)
,\label{rho1}\\
d_{\a\b}^{(\sigma)}(z)&=&-\sigma\frac{\partial}{\partial z'}\Bigg|_{z'=X^\b_\sigma}
K_{\a\b}(z,z').
\eea
The integral equations for the $f$'s are solved by formal (matrix)
Neumann-series and using this we can bring the expression for the
energy in the form
\bea
E=Lu+L\sum_\a\int_{X^\a_-}^{X^\a_+} dz\ \eps_\a^{(0)}(z)\ \rho_{a,0}(z)
-\eps^{(0)}_s(\l^h_L)+
\sum_\a\int_{X^\a_-}^{X^\a_+} dz\ \eps_\a^{(0)}(z)\ \rho_{\a,1}(z)
-\frac{\pi}{6L}(v_s+v_c),
\label{E1}
\eea
where the spin and charge velocities are
\be
v_\a=\frac{\eps'_\a(X^\a)}{2\pi\rho_{\a,0}(X^\a)}\ ,\quad
\a=c,s.
\label{velocities}
\ee
Let us denote the first two terms in \fr{E1} as $Le_{\rm
  GS}(\{X^\a_\pm\})$. We consider it as a functional of the
integration boundaries and expand it to second order around 
$\pm X^a$, i.e.
\bea
e_{\rm GS}(\{X^\a_\pm\})&=&e_{\rm GS}(\{X^\a\})
+\sum_{\b,\sigma}\left[\frac{\delta}{\delta
  X^\b_\sigma}\Biggl|_{X^\b_\sigma=\sigma X^\b}
e_{\rm GS}(\{X^\a_\pm\})\right] (X^\b_\sigma-\sigma X^\b)\nn
&&
+\frac{1}{2}\sum_{\b,\sigma,\g,\tau}\left[\frac{\delta^2}{
\delta  X^\b_\sigma\delta  X^\g_\tau}
\Biggl|_{\genfrac{}{}{0pt}{}{X^\b_\sigma=\sigma X^\b}
{X^\g_\tau=\tau X^\g}}
e_{\rm GS}(\{X^\a_\pm\})\right]
(X^\b_\sigma-\sigma X^\b)(X^\g_\tau-\tau X^\g).
\eea
We find that the linear term vanishes by virtue of the
equations \fr{kappaeps} for the dressed energies. For the quadratic
term we find after some calculations
\bea
\frac{\delta^2}{
\delta  X^\b_\sigma\delta  X^\g_\tau}
\Biggl|_{\genfrac{}{}{0pt}{}{X^\b_\sigma=\sigma X^\b}{X^\g_\tau=\tau X^\g}}
e_{\rm GS}(\{X^\a_\pm\})=\delta_{\a\b}\delta_{\sigma\tau}2\pi v_\a
\left[\rho_{\a,0}(X^\a)\right]^2.
\eea
The third and fourth terms in \fr{E1} can be simplified using the
integral equations for the dessed energies. Introducing the shift of
the hole rapidity in the finite volume by
\be
\l^h_L=\l^h+\frac{1}{L}\delta\l^h,
\label{dlambdah}
\ee
we can express the finite-size energy as
\bea
E&=&
Le_{\rm GS}(\{X^\a\})-\eps_s(\l^h)+L\pi\sum_\a  v_\a
\left\{\left[\rho_{\a,0}(X^\a)(X^\a_+-X^\a)\right]^2
+\left[\rho_{\a,0}(X^\a)(X^\a_-+X^\a)\right]^2\right\}\nn
&&-\frac{\pi}{6L}(v_s+v_c)-\frac{1}{L}\eps'_s(\l^h)\delta\l^h.
\eea
We may calculate $\delta\l^h$ from the equation
\bea
z_s(\l^h_L)=\frac{2\pi J^h}{L}&=&
\int_{Q_-}^{Q_+} dk\ \rho_{c,0}(k)\
\th\left(\frac{\l^h_L-\sin k}{u}\right)-
\int_{A_-}^{A_+} d\l\ \rho_{s,0}(\l)\
\th\left(\frac{\l^h_L-\l}{2u}\right)\nn
&&+\frac{1}{L}\int_{-Q}^{Q} dk\ \rho_{c,1}(k)\
\th\left(\frac{\l^h_L-\sin k}{u}\right)-
\frac{1}{L}\int_{-A}^{A} d\l\ \rho_{s,1}(\l)\
\th\left(\frac{\l^h_L-\l}{2u}\right)+o(L^{-1}).
\eea
Using the definition (\ref{dlambdah}) we obtain
\bea
\delta\l^h&=&-\frac{L}{2\pi\rho_{s,0}(\l^h)}
\sum_{\b,\sigma} \Psi^{(\sigma)}_\b(\l^h)
\left[X^\beta_\sigma-\sigma X^\beta\right]
\nn
&&-\frac{L}{2\pi\rho_{s,0}(\l^h)}\Biggl[
\frac{1}{L}\int_{-Q}^{Q} dk\ \rho_{c,1}(k)\
\th\left(\frac{\l^h-\sin k}{u}\right)
-
\frac{1}{L}\int_{-A}^{A} d\l\ \rho_{s,1}(\l)\
\th\left(\frac{\l^h-\l}{2u}\right)\Biggr].
\label{dlh1}
\eea
Here
\bea
\Psi^{(\sigma)}_\b(\l^h)&=&\left[
\sigma\ \rho_{\b,0}(X^\b)g_{\b,\sigma}(\l^h)+
\int_{-Q}^{Q} dk\ r^{(\sigma)}_{c\b}(k)\
\th\left(\frac{\l^h-\sin k}{u}\right)-
\int_{-A}^{A} d\l\ r^{(\sigma)}_{s\b}(\l)\
\th\left(\frac{\l^h-\l}{2u}\right)\right],
\label{dlh2}
\eea
where
$g_{c,\sigma}(\l^h)=\theta\Big(\frac{\l^h-\sigma\sin(Q)}{u}\Big)$, 
$g_{s,\sigma}(\l^h)=-\theta\Big(\frac{\l^h-\sigma A}{2u}\Big)$
and the functions $r_{\a\b}^{(\sigma)}$ fulfil the integral equations 
\be
r_{\a\b}^{(\sigma)}(z_\a)=\sigma\ \rho_{\b,0}(X^\b)\
K_{\alpha\beta}(z_\a,\sigma X^\b)
+\hat{K}_{\alpha\gamma}*r_{\gamma\b}^{(\sigma)}\Bigg|_{z_\a}.
\label{dlh3}
\ee
\subsection{Relating $X^\a_\sigma-\sigma X^\a$ to Quantum Numbers}
In the next step we want to express the deviations of the integration
boundaries from their thermodynamic values through appropriate quantum
numbers. We define $N_{c,s}$ by
\bea
n_c&=&\frac{N_c}{L}=\frac{I_+-I_-}{L}=\int_{Q_-}^{Q_+}dk\ \rho_c(k)\ ,\nn
n_s&=&\frac{N_s}{L}=\frac{J_+-J_-}{L}=\int_{A_-}^{A_+}d\l\ \rho_s(\l)\ .
\label{ncs}
\eea
We note that the number of down spins is $N_s-1$ rather than $N_s$ as we
have one ``deep'' hole in the distribution of $\l$'s. The other
quantities we want to use are
\bea
2D_c&=&I_++I_-=\frac{L}{2\pi}\left[z_c(Q_+)+z_c(Q_-)\right]\ ,\nn
2D_s&=&J_++J_-=\frac{L}{2\pi}\left[z_s(A_+)+z_s(A_-)\right].
\eea
Using the integral equations \fr{count} for the counting functions
$z_{c,s}$ we can rewrite these as
\bea
\label{dcs}
\frac{2D_s}{L}&=&2d_s=\int_{-\infty}^{A_-}d\l \rho_s(\l)
-\int_{A_+}^{\infty}d\l \rho_s(\l),\nn
\frac{2D_c}{L}&=&2d_c=\int_{-\pi}^{Q_-}dk\ \rho_c(k)
-\int_{Q_+}^{\pi}dk\ \rho_c(k)-\frac{1}{\pi}\int_{A_-}^{A_+}d\l\
\th\left(\frac{\l}{u}\right)\ \rho_s(\l)+\frac{1}{\pi L}
\th\left(\frac{\l^h}{u}\right).
\eea
We have to order $1/L$
\bea
n_\a&=&\int_{X^\a_-}^{X^\a_+}dz\ \rho_{\a,0}(z)
+\frac{1}{L}\int_{-X^\a}^{X^\a}dz\ \rho_{\a,1}(z).
\eea
The second term no longer depends on $X^\a_\sigma$ and is denoted by
\bea
N_c^{\rm imp}&=&\int_{-Q}^{Q}dk\ \rho_{c,1}(k)\ ,\nn
N_s^{\rm imp}&=&\int_{-A}^{A}dk\ \rho_{s,1}(\l)\ .
\label{Nimp}
\eea
The variation of the integration boundaries $X^\a_\sigma$ with $n_\b$
is now easily calculated to leading order in $L^{-1}$
\be
\frac{\partial X^\a_\pm}{\partial n_\b}=
\pm\frac{1}{2}\frac{Z^{-1}_{\a\b}}{\rho_{\a,0}(X^\a)}.
\label{xn}
\ee
Here the dressed charge matrix $Z$
\be
Z=
\begin{pmatrix}
\xi_{cc}(Q) & \xi_{cs}(A)\cr
\xi_{sc}(Q) & \xi_{ss}(A)
\end{pmatrix}\ ,
\label{Z}
\ee
is given by $Z_{\a\b}=\xi_{\a\b}(X^\b)$, where $\xi_{\a\b}$ fulfil the
set of coupled integral equations 
\be
\xi_{\a\b}(z_\b)=\delta_{\a\b}+\xi_{\a\g}*\widehat{K}_{\g\b}\Bigg|_{z_\b}.
\label{xi}
\ee
Similarly we have
\bea
2d_s&=&\int_{-\infty}^{A_-}d\l \rho_{s,0}(\l)
-\int_{A_+}^{\infty}d\l \rho_{s,0}(\l)+\frac{2}{L}D_s^{\rm imp},\nn
2d_c&=&\int_{-\pi}^{Q_-}dk\ \rho_{c,0}(k)
-\int_{Q_+}^{\pi}dk\ \rho_{c,0}(k)-\frac{1}{\pi}\int_{A_-}^{A_+}d\l\
\th\left(\frac{\l}{u}\right)\ \rho_{s,0}(\l)+\frac{2}{L}D_c^{\rm imp}\ ,
\eea
where
\bea
2D_s^{\rm imp}&=&\int_{-\infty}^{-A}d\l \rho_{s,1}(\l)
-\int_{A}^{\infty}d\l \rho_{s,1}(\l)\ ,\nn
2D_c^{\rm imp}&=&\int_{-\pi}^{-Q}dk\ \rho_{c,1}(k)
-\int_{Q}^{\pi}dk\ \rho_{c,1}(k)-\frac{1}{\pi}\int_{-A}^{A}d\l\
\th\left(\frac{\l}{u}\right)\ \rho_{s,1}(\l)
+\frac{1}{\pi}\th\left(\frac{\l^h}{u}\right).
\label{Dimp}
\eea
These allows us calculate the dependence of the integration boundaries
on $d_\a$
\be
\frac{\partial X^\a_\pm}{\partial d_\b}=
\frac{Z_{\b\a}}{\rho_{\a,0}(X^\a)}.
\label{xd}
\ee
Combining \fr{xn} with \fr{xd} we obtain the described relationship
between the change in integration boundaries and the quantum numbers
$n_\a$, $d_\a$
\bea
X^\a_\pm\mp
X^\a&=&\pm\frac{1}{2}\frac{Z^{-1}_{\a\b}}{\rho_{\a,0}(X^\a)}
\left[\Delta n_\b-\frac{1}{L}N_\b^{\rm imp}\right]
+\frac{Z^T_{\a\b}}{\rho_{\a,0}(X^\a)}\left[\Delta d_\b-\frac{1}{L}D_\b^{\rm imp}\right]\ .
\eea
Here $\Delta n_{c,s}$ and $\Delta d_{c,s}$ are the changes in the
quantum numbers \fr{ncs}, \fr{dcs} compared to the ground state.
\subsection{Result for the Finite Size Energy}
Putting everything together we then arrive at
\bea
E&=&Le_{\rm GS}(\{X^\a\})-\eps_s(\l^h)\nn
&&+\frac{1}{L}\left\{
-\frac{\pi}{6}(v_s+v_c)
+2\pi\left[
\frac{1}{4}\Delta \tilde{N}_\gamma (Z^T)^{-1}_{\gamma\a}v_\a Z^{-1}_{\a\beta}
\Delta \tilde{N}_\beta+
\tilde{D}_\gamma Z_{\gamma\a}v_\a Z^T_{\a\beta}
\tilde{D}_\beta\right]-\eps'_s(\l^h)\delta\l^h\right\}.
\label{FSE}
\eea
Here we have defined
\bea
\Delta \tilde{N}_\a&=&\Delta N_\a-N^{\rm imp}_\a\ ,\nn
\tilde{D}_\a&=& D_\a-D^{\rm imp}_\a\ .
\label{tildeND}
\eea
In writing \fr{tildeND} we have used that for the ground state $D_\a=0$.
The first term in \fr{FSE} is the ground state energy, the second term
the excitation energy in the thermodynamic limit and the second line
gives the $L^{-1}$ corrections. For our case we have
\be
N_c=I_+-I_-=N_{\rm GS}-1\ ,\quad N_s=J_+-J_-=M_{\rm GS}\ ,\quad
D_c=\frac{I_++I_-}{2}=0\ ,\quad
D_s=\frac{J_++J_-}{2}=0\ .
\ee
For the ground state we have
\be
N_c=N_{\rm GS}\ ,\quad N_s=M_{\rm GS}\ ,
\ee
which gives
\be
\Delta N_c=-1\ ,\quad \Delta N_s=0.
\label{DcsNcs}
\ee 
\subsection{Relation of $N_\a^{\rm imp}$ to the spectrum}
The quantities $N_\a^{\rm imp}$ are given in terms of the solutions to
the coupled integral equations \fr{inteq}. It is possible to relate
them to properties of the dispersions of the elementary excitations as
follows. 
The integral equations for $\rho_{\alpha,1}$ are formally solved by 
\bea
\rho_{\alpha,1}=(1-\hat{K})^{-1}_{\a\b}*\rho_{\beta,1}^{(0)}
=-(1-\hat{K})^{-1}_{\a\b}*K_{\beta s}.
\eea
Substituting this into the equations for $N^{\rm imp}_\a$ we obtain
\bea
N^{\rm imp}_c&=&-\int_{-Q}^Qdk\ (1-\hat{K})^{-1}_{c\b}*K_{\beta
  s}=-\int_{-Q}^Qdk\ (1-\hat{K})^{-1}_{cs}(k,\l^h)\ ,\nn
N^{\rm imp}_s&=&-\int_{-A}^Ad\l\ (1-\hat{K})^{-1}_{s\b}*K_{\beta s}
=1-\int_{-A}^Ad\l\ (1-\hat{K})^{-1}_{ss}(\l,\l^h).
\eea
Here we have used e.g.
\be
(1-\hat{K})^{-1}_{c\b}*K_{\beta  s}=
(1-\hat{K})^{-1}_{c\b}*(\hat{K}-1+1)_{\beta  s}=
(1-\hat{K})^{-1}_{cs}.
\ee
On the other hand, by differentiating the integral equations
\fr{kappaeps} for the dressed energies we obtain
\be
\frac{\partial\eps_\a}{\partial \mu}=
\frac{\partial\eps_\b}{\partial \mu}*\hat{K}_{\b\a}
-\delta_{\a c},
\label{depsdmu}
\ee
which are solved by
\be
\frac{\partial\eps_\a}{\partial \mu}=-\int_{-Q}^Qdk\ 
(1-\hat{K})^{-1}_{c \a}(k,z_\a).
\ee
This gives us our first relation
\be
N_c^{\rm imp}=\frac{\partial\eps_s(\l^h)}{\partial \mu}.
\label{nceps}
\ee
By comparing \fr{depsdmu} to \fr{xi} we observe that
\be
\frac{\partial\eps_\a}{\partial \mu}(z_\a)=-\xi_{c\a}(z_\a),
\ee
which in conjunction with \fr{nceps} allows us to relate $N_c^{\rm
  imp}$ to the dressed charge matrix as
$N_c^{\rm imp}=-\xi_{cs}(\l^h)$. This generalizes the analogous
relation for the spin-1/2 XXZ chain found in Ref.[\onlinecite{Affleck2}].
A second relation is obtained by considering
\be
\frac{\partial\eps_\a}{\partial B}=\frac{\partial\eps_\b}{\partial B}*
\hat{K}_{\b\a}+2\delta_{\a s}-\delta_{\a c}.
\label{epsB}
\ee
Comparing this to \fr{xi} and using the linearity of the integral
equation we observe that
\be
\frac{\partial\eps_\a}{\partial B}=2\xi_{s\a}-\xi_{c\a}.
\ee
The formal solution of \fr{epsB} is
\be
\frac{\partial\eps_\a}{\partial B}=
\frac{\partial\eps_\a}{\partial
  \mu}+2\int_{-A}^Ad\l\ (1-\hat{K})^{-1}_{s\a}.
\ee
Our second relation is then
\be
N_s^{\rm imp}=1-\frac{1}{2}\left[
\frac{\partial\eps_s(\l^h)}{\partial B}-
\frac{\partial\eps_s(\l^h)}{\partial \mu}\right]=1-\xi_{ss}(\l^h).
\ee
\subsection{Simplification for zero Magnetic Field}
In the absence of a magnetic field we have $A=\infty$ which allows us
to simplify our results for the finite-size corrections to the energy
\fr{FSE}. The dressed charge matrix takes the form \cite{book}
\bea
Z=\begin{pmatrix}
\xi & 0\cr
\frac{\xi}{2} & \frac{1}{\sqrt{2}}
\end{pmatrix}\ ,
\label{xiH0}
\eea
where $\xi=\xi(Q)$ is obtained from the solution of the integral equation
\be
\xi(k)=1+\int_{-Q}^Qdk'\ \cos(k')\ R(\sin(k)-\sin(k'))\ \xi(k').
\label{xiB0}
\ee
Here the function $R(x)$ is
\be
R(x)=\int_{-\infty}^\infty \frac{d\omega}{2\pi}\frac{e^{i\omega x}}
{1+\exp(2u|\omega|)},
\label{rofx}
\ee
and can be expressed in terms of the Digamma function. The integral
equations for the dressed energies and root densities simplify to \cite{book}
\bea
\eps_c(k)&=&-2\cos(k)-\mu-2u+\int_{-Q}^Qdk'\ \cos(k')\ R(\sin(k)-\sin(k'))\
\eps_c(k')\ ,\nn
\eps_s(\l)&=&\int_{-Q}^Qdk\cos(k)\ s(\l-\sin(k))\ \eps_c(k)\ ,\nn
\rho_c(k)&=&\frac{1}{2\pi}+\cos(k)\int_{-Q}^Qdk'\ R(\sin(k)-\sin(k'))\
\rho_c(k')\ ,\nn
\rho_s(\l)&=&\int_{-Q}^Qdk\ s(\l-\sin(k))\ \rho_c(k)\ ,
\label{rhoepsB0}
\eea
where
\be
s(x)=\frac{1}{4u\cosh\Big(\frac{\pi x}{2u}\Big)}.
\label{sofx}
\ee
The finite-size energy can be expressed as
\bea
E&=&Le_{\rm GS}(\{X^\a\})-\eps_s(\l^h)
-\frac{\pi}{6L}(v_s+v_c)-\frac{1}{L}\eps_s'(\l^h)\delta\l^h\nn
&+&\frac{2\pi  v_c}{L}
\left[\frac{(\Delta N_c-N_c^{\rm imp})^2}{4\xi^2}
+\xi^2\left({D}_c-D^{\rm imp}_c+\frac{{D}_s}{2}\right)^2\right]
+\frac{2\pi
  v_s}{L}
\left[\frac{1}{2}\left(\Delta {N}_s-\frac{\Delta {N}_c}{2}-\frac{1}{2}\right)^2
+\frac{{D}_s^2}{2}\right], 
\label{EofLh0}
\eea
where
\bea
N^{\rm imp}_c&=&2N_s^{\rm imp}-1=\int_{-Q}^Qdk \rho_{c,1}(k)\ ,\\
D^{\rm imp}_s&=&0\ ,\\
2D^{\rm imp}_c&=&
\int_{Q}^{\pi}dk\left[ \rho_{c,1}(-k)- \rho_{c,1}(k)\right]
+\frac{1}{\pi}\theta\left(\tanh\left(\frac{\pi\l^h}{4u}\right)\right)\nn
&&-\frac{1}{\pi}\int_{-Q}^Qdk\ \rho_{c,1}(k)\
i\ln\left[
\frac{\Gamma\left(\frac{1}{2}+i\frac{\sin k}{4u}\right)}
{\Gamma\left(\frac{1}{2}-i\frac{\sin k}{4u}\right)}
\frac{\Gamma\left(1-i\frac{\sin k}{4u}\right)}
{\Gamma\left(1+i\frac{\sin k}{4u}\right)}\right].
\label{H0NDimp}
\eea
Here $\rho_{c,1}$ is the solution to the integral equation
\bea
\rho_{c,1}(k)=-\frac{\cos(k)}{4u\cosh\left(\frac{\pi(\l^h-\sin
      k)}{2u}\right)}
+\cos(k)\int_{-Q}^Q dk'\ R(\sin k-\sin k')\ \rho_{c,1}(k').
\label{rhoc1}
\eea
The fact that $D_s^{\rm imp}=0$ is established in Appendix \ref{app:H0}.
The relation \fr{nceps} of $N_c^{\rm imp}$ to the dressed energy of spin
excitations remains valid for zero magnetic field, i.e. we have that
$N_c^{\rm imp}=\frac{\partial\eps_s(\Lambda^h)}{\partial\mu}$.
The quantum numbers $\Delta N_\a$, $D_\a$ for the holon-spinon
excitation in zero magnetic field are 
\bea
D_c= D_s=0\ ,\quad
\Delta N_c&=&-1\ ,\quad
\Delta N_s=0.
\eea
This shows that there are no $L^{-1}$ corrections to the excitation
energy in the spin sector. In terms of the field theory picture (see
section \ref{sec:FT})of a deep hole interacting with the gapless
Luttinger liquid degrees of freedom this shows that the interaction of
the deep hole with the gapless spin sector is indeed irrelevant.
Due to the presence of the marginally irrelevant interaction of spin
currents we expect logarithmic corrections, but these are beyond the
accuracy of our finite-size calculation.

\begin{figure}[ht]
\begin{center}
(a)\epsfxsize=0.45\textwidth
\epsfbox{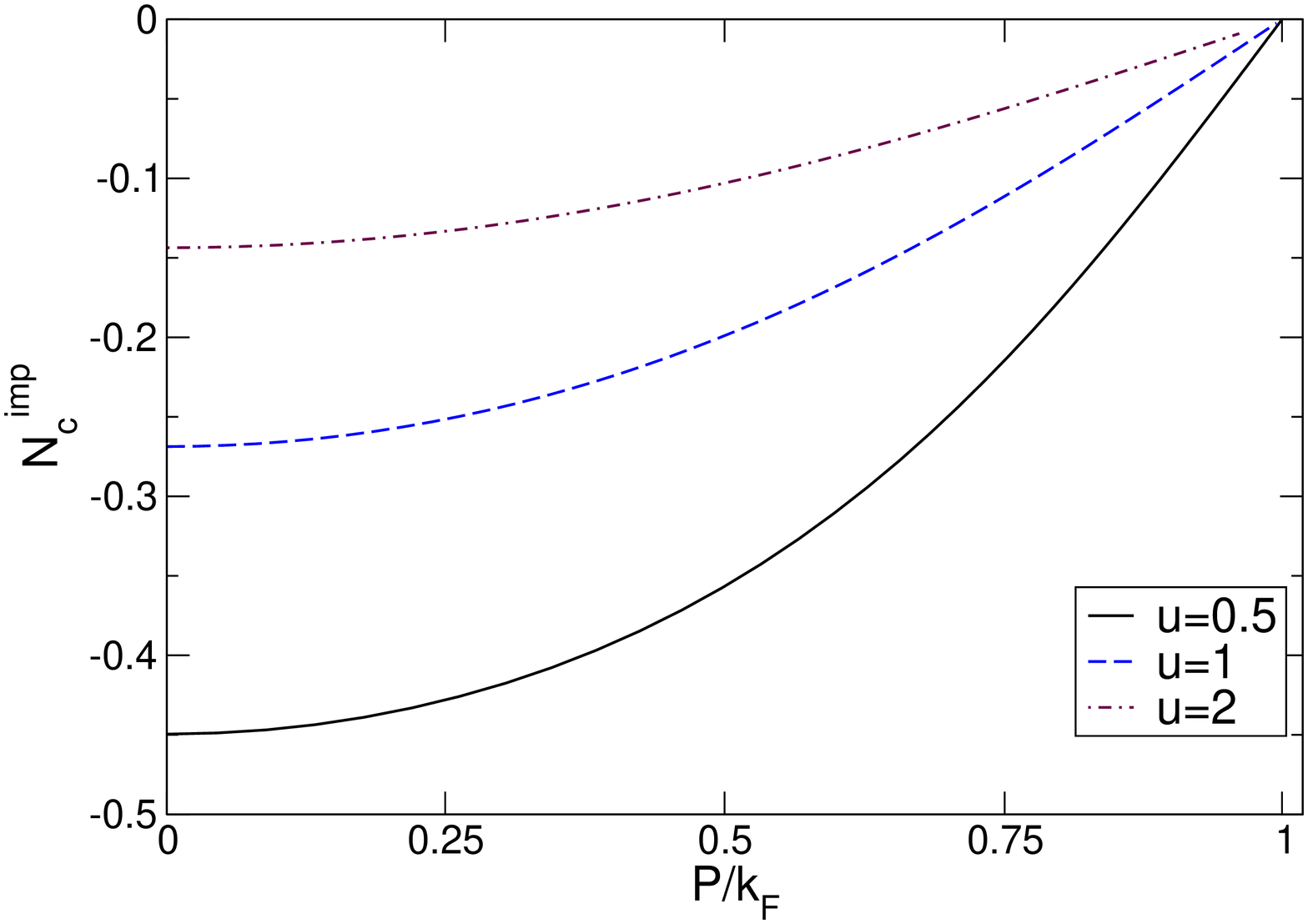}\qquad
(b)\epsfxsize=0.45\textwidth
\epsfbox{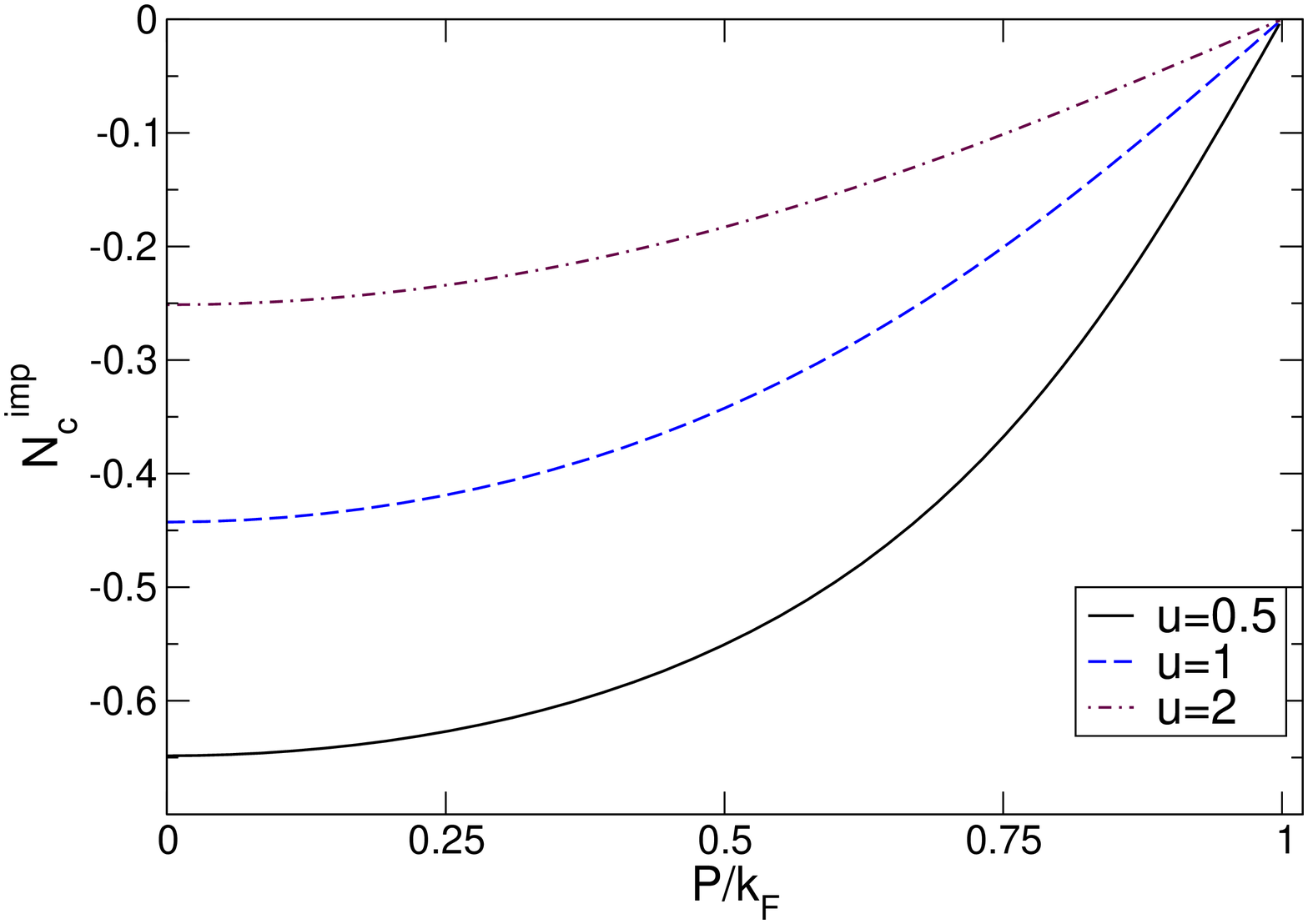}
\end{center}
\caption{$N_c^{\rm imp}$ as a function of momentum for $u=0.5,1,2$ and 
densities (a) $n_c=0.2$ and (b) $n_c=0.5$.}
\label{fig:ncimp}
\end{figure}

\begin{figure}[ht]
\begin{center}
(a)\epsfxsize=0.45\textwidth
\epsfbox{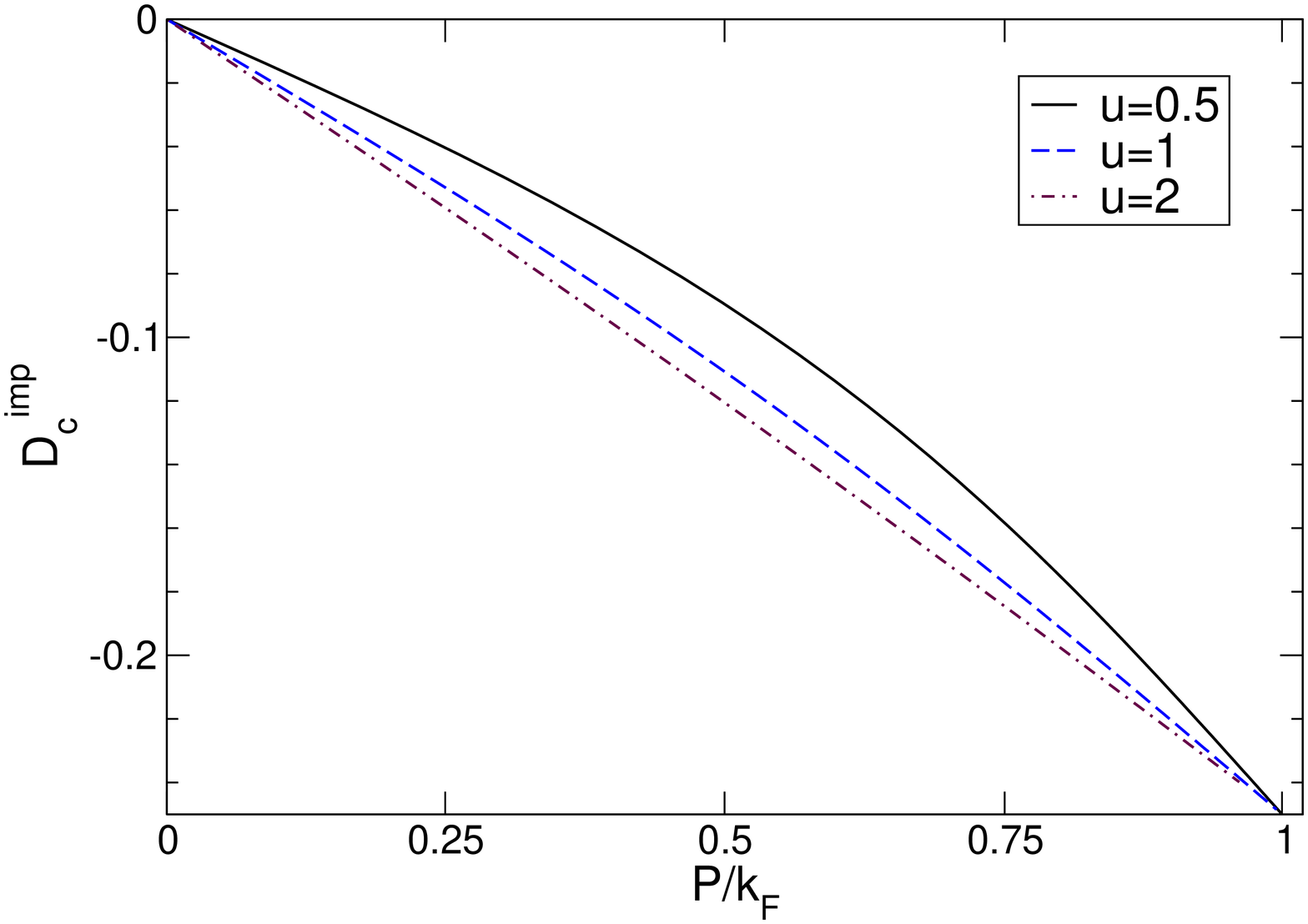}\qquad
(b)\epsfxsize=0.45\textwidth
\epsfbox{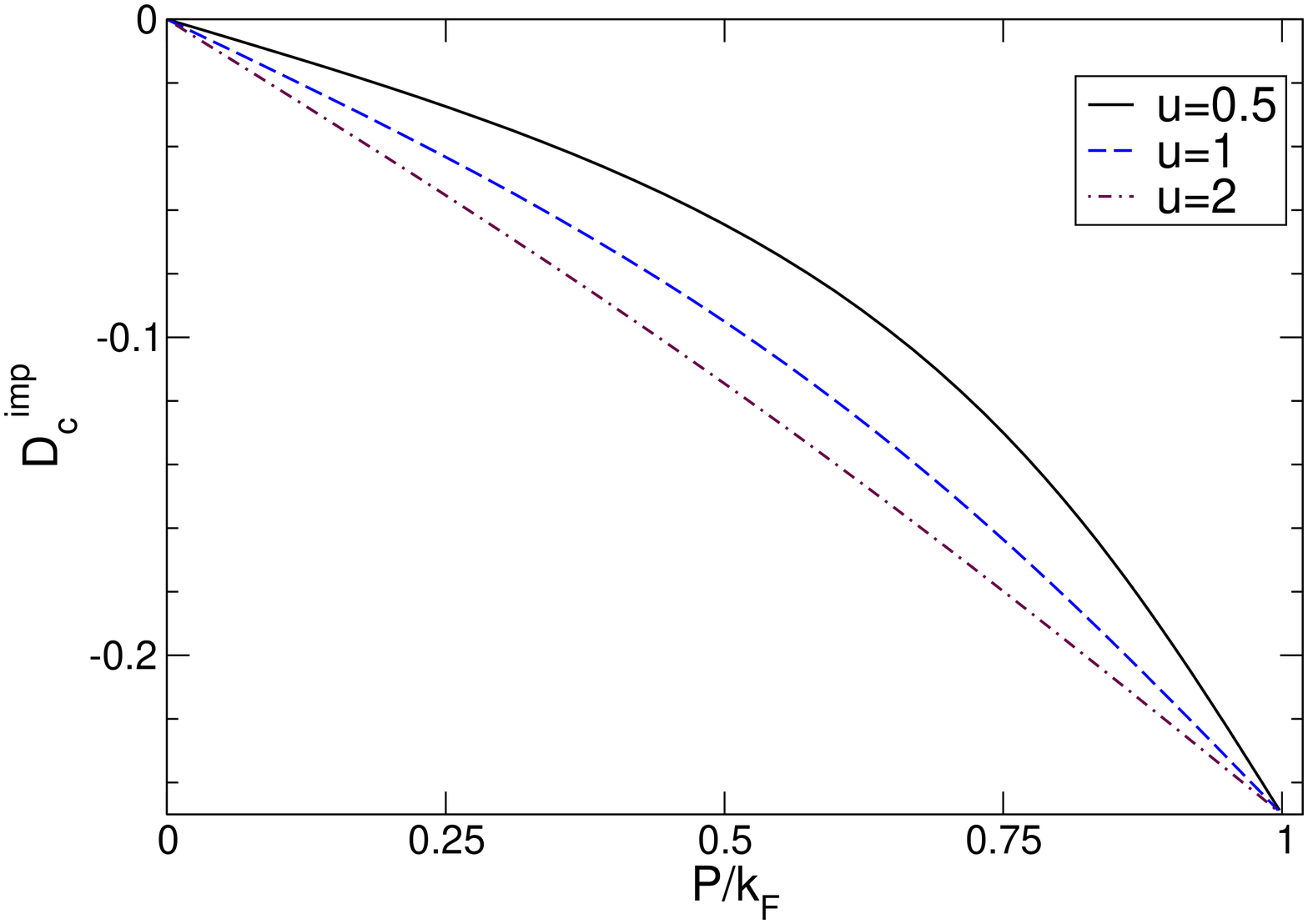}
\end{center}
\caption{$D_c^{\rm imp}$ as a function of momentum for $u=0.5,1,2$ and 
densities (a) $n_c=0.2$ and (b) $n_c=0.5$.}
\label{fig:dcimp}
\end{figure}

Explicit values for the quantities $N_c^{\rm imp}$ and $D_c^{\rm imp}$
are readily obtained from a numerical solution of the linear integral
equation \fr{rhoc1}. We present results for several values of $u$ and
two band fillings in Figs\ref{fig:ncimp} and \ref{fig:dcimp}.

\subsubsection{Large-$u$ Limit}
In the strong interaction limit $u\gg 1$ we can solve the integral
equation \fr{rhoc1} by iteration and obtain explicit results. To
leading order we find
\be
D_c^{\rm imp}\approx \frac{p_s(\l^h)}{2\pi n_c}=-\frac{P_{hs}}{4k_F}\ ,\qquad
N_c^{\rm imp}\approx -\frac{\sin(Q)}{2u\cosh(\pi\l^h/2u)}.
\label{largeU}
\ee

\section{Threshold at $k_F<P<2k_F$ for the Holon - 3 Spinon Excitation}
\label{sec:thres2}
\begin{figure}[ht]
\begin{center}
\epsfxsize=0.6\textwidth
\epsfbox{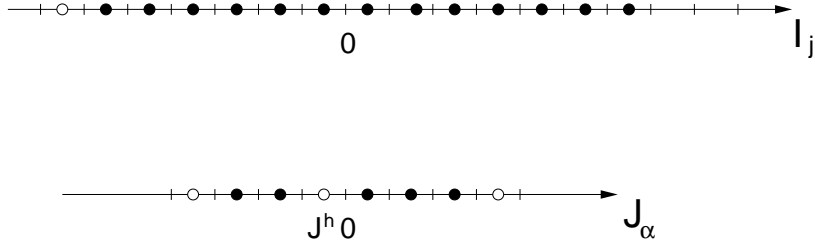}
\end{center}
\caption{Distribution of integers for the threshold of the holon - 3
  spinon excitation. The hole in the distribution of $I_j$ is at
$-(N_{\rm GS}-1)/2$ and the three holes in the distribution of $J_\a$
  occur at $\pm M_{\rm GS}/2$ and $J^h$ respectively.} 
\label{fig:threshsss}
\end{figure}

We now want to determine the finite-size corrections to the energy of
the holon - 3 spinon excitation along the low-energy threshold in the
momentum range $k_F<P<2k_F$. The threshold of this excitation is
described by the Bethe Ansatz equations \fr{BAE} with $N=N_{\rm
  GS}-1$, $M=M_{\rm GS}-2$ and half-odd integer numbers
\bea
I_j&=&-\frac{N_{\rm GS}-1}{2}+j\ ,\quad 1\leq j\leq N_{\rm GS}-1\ ,\nn
J_\a&=&\begin{cases}
-\frac{M_{\rm GS}}{2}+\a & \text{if}\ 1\leq \a<\frac{M_{\rm GS}}{2}+J^h\cr
-\frac{M_{\rm GS}}{2}+\a+1 & \text{if}\ \frac{M_{\rm GS}}{2}+J^h\leq
\a\leq M_{\rm GS}-2\cr
\end{cases}.
\eea
The corresponding distributions of half-odd integers are shown in
Fig.\ref{fig:threshsss}.
The calculation on the finite-size corrections to the energy now
proceeds just like for the holon-spinon excitation. The result is
given by \fr{FSE}, \fr{tildeND}, \fr{Nimp}, \fr{Dimp} where now
\be
N_c=I_+-I_-=N_{\rm GS}-1\ ,\quad N_s=J_+-J_-=M_{\rm GS}-1\ ,\quad
D_c=\frac{I_++I_-}{2}=\frac{1}{2}\ ,\quad
D_s=\frac{J_++J_-}{2}=0\ .
\ee
Recalling that in the ground state the $N_{c,s}$ ``quantum numbers'' are
\be
N_c=N_{\rm GS}\ ,\quad N_s=M_{\rm GS}\ ,
\ee
we conclude that for the holon - 3 spinon excitation we have
\be
\Delta N_c=-1\ ,\quad \Delta N_s=-1\ ,\quad D_c=\frac{1}{2}\ ,\quad D_s=0.
\label{DcsNcs_hsss}
\ee 
The reduction in the zero magnetic field case is completely analogous
to the holon-spinon excitation. Hence the energy is again given by
\fr{EofLh0}, \fr{H0NDimp}, \fr{rhoc1}, but the ``quantum numbers''
$N_{c,s}$, $D_{c,s}$ are now given by \fr{DcsNcs_hsss}.
\section{Field Theory}
\label{sec:FT}
We now relate our results to the field theory treatment of
Ref.~\onlinecite{SIG} for the threshold singularity problem. There it
was shown that a high-energy excitation in a spinful Luttinger liquid
can be mapped to a mobile impurity in a Luttinger liquid. The
corresponding Hamiltonian is $H=\sum_{\a=c,s} H_\a+H_d+H_{\rm int}$, where
\bea
H_\a&=&\frac{v_\a}{2\pi}\int dx\left[
\frac{1}{K_\a}\left(\frac{\partial \Phi_\a}{\partial x}\right)^2
+K_\a\left(\frac{\partial \Theta_\a}{\partial x}\right)^2\right]\ ,\nn
H_d&=&\int dx\ d^\dagger(x)\left[\varepsilon_s(P)-iu\partial_x\right]d(x)\ ,\nn
H_{\rm int}&=&\int dx\ \left[\frac{V_R-V_L}{2\pi}\partial_x\Theta_c
-\frac{V_R+V_L}{2\pi}\partial_x\Phi_c \right]d(x)d^\dagger(x).
\label{HMI}
\eea
Here the Bose fields $\Phi_\a$ and the dual fields
$\Theta_\a$ fulfil the commutation relations \fr{comm}, $d(x)$ and
$d^\dagger$ are annihilation and creation operators of the mobile
impurity, which carries momentum $P$ and travels at velocity $u$.
The parameters $V_{R,L}$ characterize the interaction of the impurity
with the low energy charge degrees of freedom. The parameters of
$H_{c,s}$ and $H_d$ in \fr{HMI} are readily identified with quantities
obtained from the Bethe Ansatz. The spin and charge velocities
$v_{s,c}$ are given by \fr{velocities} and the Luttinger parameters are
\be
K_s=1\ ,\quad K_c=\frac{\xi^2}{2},
\ee
where $\xi$ is given by \fr{xiB0}. The velocity of the impurity is
expressed in terms of the solutions to the integral equations
\fr{rhoepsB0} as 
\be
u=\frac{\eps'_s(\Lambda^h)}{2\pi \rho_s(\Lambda^h)}.
\ee
The ``chemical potential'' of the impurity is $\eps_s(\Lambda^h)$,
where the position $\Lambda^h$ of the hole is fixed by the requirement
\be
P_{hs}(\Lambda^h)=P.
\ee
The parameters $V_{R,L}$ entering the expression for $H_{\rm int}$ are
deterined as follows. Following Refs~\onlinecite{Affleck2,SIG} we
remove the interaction term $H_{\rm int}$ through a unitary
transformation on the fields
\bea
U=\exp\left(-i\int dx\Big[ 
\sqrt{K_c}\frac{\Delta\delta_{+,c}-\Delta\delta_{-,c}}{2\pi}\Theta_c(x)
-\frac{\Delta\delta_{+,c}+\Delta\delta_{-,c}}{2\pi\sqrt{K_c}}\Phi_c(x)
\Big]d(x)d^\dagger(x)\right),
\label{udagger}
\eea
where
\be
-(V_L\mp V_R)K_c^{\mp\frac{1}{2}}=(v_c+u)\Delta\delta_{-,c}\pm(v_c-u)\Delta\delta_{+,c}.
\ee
In the resulting Hamiltonian the impurity no longer interacts
explicitly with the charge part of Luttinger liquid, but it does
affect the boundary conditions of the charge boson. In particular we
find that
\bea
\partial_x\widehat{\Phi}_c&=&U^\dagger\partial_x\Phi_cU
=\partial_x\Phi_c-\frac{\sqrt{K_c}}{2}
\Big(\Delta\delta_{+,c}-\Delta\delta_{-,c}\Big)dd^\dagger\ ,\nn
\partial_x\widehat{\Theta}_c&=&
U^\dagger\partial_x\Theta_cU=\partial_x\Theta_c+\frac{1}{2\sqrt{K_c}}
\Big(\Delta\delta_{+,c}+\Delta\delta_{-,c}\Big)dd^\dagger\ .
\label{shifts}
\eea
Taking the expectation value of \fr{shifts} with respect to a state
with a high-energy hole we find that
\bea
\int dx\ \langle\partial_x\Theta_c\rangle&=&
\int dx\ \langle\partial_x\widehat{\Theta}_c\rangle-
\frac{1}{2\sqrt{K_c}}
\Big(\Delta\delta_{+,c}+\Delta\delta_{-,c}\Big)\ ,\nn
\int dx\ \langle\partial_x\Phi_c\rangle&=&
\int dx\ \langle\partial_x\widehat{\Phi}_c\rangle
+\frac{\sqrt{K_c}}{2}
\Big(\Delta\delta_{+,c}-\Delta\delta_{-,c}\Big).
\label{shifts2}
\eea
Denoting by $\rho_{c,R}$ and $\rho_{c,L}$ the charge densities at the
right and left Fermi wave numbers respectively we have that the
numbers $\Delta N$ and $D$ of low-energy charge and current
excitations is
\bea
\Delta N&=&\int dx\sqrt{2}\left[\rho_{c,R}+\rho_{c,L}\right]
=-\frac{\sqrt{2}}{\pi}\int dx\ \langle \partial_x\Phi_c\rangle\ ,\nn
4D&=&\sqrt{2}\int dx\left[\rho_{c,R}-\rho_{c,L}\right]
=\frac{\sqrt{2}}{\pi}\int dx\ \langle \partial_x\Theta_c\rangle\ .
\eea
We are now in a position to identify the additional contributions in
\fr{shifts2} that arise from the interaction of the Luttinger liquid
with the mobile impurity with the quantities $N_c^{\rm imp}$ and
$D_c^{\rm imp}$.
\subsection{Holon-Spinon Exitation}
\begin{figure}[ht]
\begin{center}
\epsfxsize=0.5\textwidth
\epsfbox{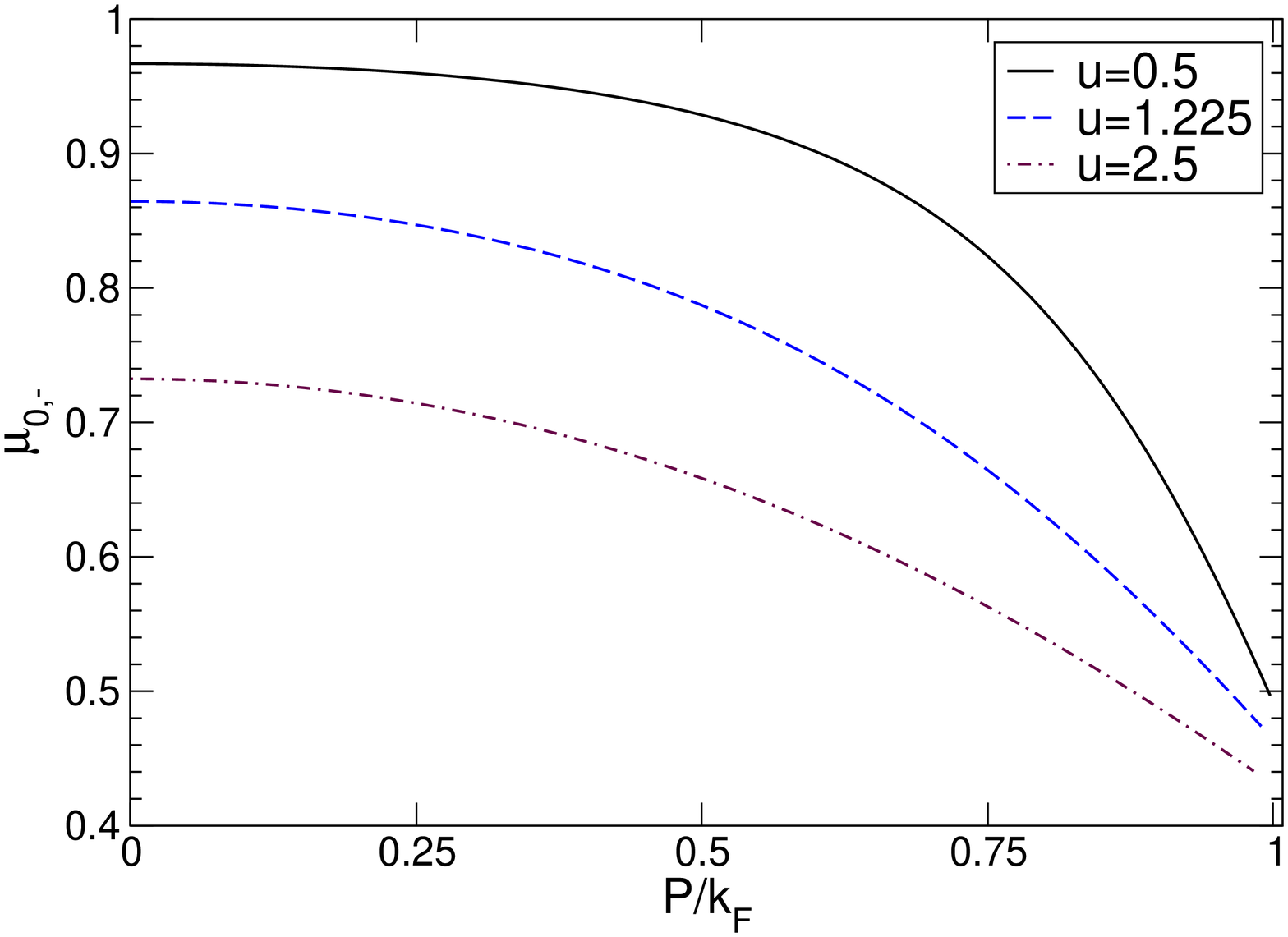}
\end{center}
\caption{Threshold singularity exponent $\mu_{0,-}$ as a function of
  momentum for $n_c=0.6$.}
\label{fig:mu0-}
\end{figure}

Taking into acount that for the holon-spinon excitation we have
\be
-\frac{\sqrt{2}}{\pi}\int dx\ 
\langle\partial_x{\Phi}_c\rangle=-1\ ,\quad
\frac{\sqrt{2}}{\pi}\int dx\ \langle\partial_x{\Theta}_c\rangle=-1\ ,
\ee
we conclude that the quantities $N_c^{\rm imp}$ and $2D_c^{\rm imp}$
are related to the phases $\Delta\delta_{\pm,c}$ by
\bea
N^{\rm imp}_c&=&-\sqrt{2K_c}\ \frac{\Delta\delta_{+,c}-\Delta\delta_{-,c}}{2\pi}\ ,\nn
2D_c^{\rm imp}&=&\frac{1}{2}-\frac{1}{\sqrt{2K_c}}\
\frac{\Delta\delta_{+,c}+\Delta\delta_{-,c}}{2\pi}.
\eea
In Ref.~\onlinecite{SIG} it was shown that in the momentum range
$|P|<k_F$ the single particle spectral
function exhibits a threshold singularity of the form
\be
A(\omega,P)\propto \left(\omega-\eps_s(P)\right)^{-\mu_{0,-}}\ ,
\ee
where the exponent is expressed in terms of the phase-shifts
$\Delta\delta_{\pm,c}$ by
\bea
\mu_{0,-}&=&1-\frac{1}{2}\left[-\sqrt{\frac{K_c}{2}}
+\frac{\Delta\delta_{+,c}+\Delta\delta_{-,c}}{2\pi}\right]^2
-\frac{1}{2}\left[\frac{1}{\sqrt{2K_c}}
-\frac{\Delta\delta_{+,c}-\Delta\delta_{-,c}}{2\pi}\right]^2\nn
&=&1-K_c \Big(2D_c^{\rm imp}\Big)^2-\frac{1}{4K_c}\Big(1+N_c^{\rm imp}\Big)^2.
\label{exponent}
\eea
We note that \fr{exponent} differs from the Luttinger liquid result
\cite{Lutt,FSspectrum} 
\be
\mu_{LL}=1-\frac{K_c}{4}-\frac{1}{4K_c}.
\ee
However, in the zero energy limit $P\to k_F$ the Luttinger liquid
result is recovered. Explicit results for $\mu_{0,-}$ are
obtained by solving the integral equation \fr{rhoc1} numerically. In
Fig.\ref{fig:mu0-} we present results for density $n_c=0.6$ and three
different values of $u$. 

\subsubsection{Large-$u$ Limit}
For $u\gg 1$ we have $K_c\approx\frac{1}{2}$ and using \fr{largeU} we
find
\be
\mu_{0,-}\approx\frac{1}{2}-\frac{1}{8}\left[\frac{P}{k_F}\right]^2.
\label{expinf}
\ee
This suggests that the singularity is only weakly momentum dependent
and is close to a square root. The result \fr{expinf} agrees with the
exponent computed by exploiting the factorization of the wave
function into spin and charge parts \cite{ogata,inftyU,carmelo}. 

\subsubsection{Comparison to DMRG Results}
In Ref.\onlinecite{jeckel} the single particle spectral function for
parameter values $u=1.225$ and $n_c=0.6$ was computed by the dynamical
density matrix renormalization group method \cite{jeckel2}. The
following two exponents were reported based on a scaling analysis of
the low-energy peak heights
\be
\mu_{0,-}\approx 0.86 \ {\rm for}\ P=0\ ,\qquad
\mu_{0,-}\approx 0.78 \ {\rm for}\ P=\frac{\pi}{10}.
\ee
The Luttinger liquid parameter and Luttinger liquid threshold exponent
are $K_c=0.6851$ and $\mu_{LL}=0.4638$ respectively. Our results are
\be
\mu_{0,-}\approx 0.864 \ {\rm for}\ P=0\ ,\qquad
\mu_{0,-}\approx 0.832 \ {\rm for}\ P=\frac{\pi}{10}.
\ee
The agreement for $P=0$ is excellent, but the $P=\frac{\pi}{10}$ values
are very different. One possible explanation for this discrepancy is as
follows. The frequency range over which the power-law \fr{exponent}
holds can a priori be quite small and furthermore is expected to
narrow as $P$ increases from zero to $k_F$. Extracting the threshold
exponent from the scaling of the peak height could then require very
small values of the parameter $\eta$ in Ref.~\onlinecite{jeckel}.

\subsection{Holon - 3 Spinon Exitation}
The threshold of the holon - 3 spinon excitation can also be analyzed
in terms of a mobile impurity field theory model \cite{SIG,work4}.
The threshold behaviour in the range $k_F<P<3k_F$ is no longer
singular but describes a power law ``shoulder''. The exponent can be
expressed in terms of the quantities $D_c^{\rm imp}$, $N_c^{\rm imp}$
derived above.

\section{Electron Gas}
\label{sec:egas}
The analysis for the electron gas \cite{YG} is completely analogous to
the one for the Hubbard model. The Hamiltonian is
\bea
H=-\sum_{j=1}^N\frac{\partial^2}{\partial
  x_j^2}+4u\sum_{i<j}\delta(x_i-x_j)-\mu N+B(2M-N).
\eea
The Bethe Ansatz equations read \cite{YG,Taka}
\begin{eqnarray} \label{t1EG}
&&     k_j L  =  2 \pi I_j - \sum_{\alpha = 1}^{M}
                 \theta \left(
		 \frac{k_j - \Lambda_\alpha}{u} \right),\quad
               j=1,\ldots,N\ ,
		 \\ \label{t2EG}
&&     \sum_{j=1}^{N} \theta \left(
		 \frac{\Lambda_\alpha - k_j}{u} \right)  = 
		 2 \pi J_\alpha+
		 \sum_{\beta = 1}^{M}
		 \theta \left(
		 \frac{\Lambda_\alpha - \Lambda_\beta}{2u} \right)\ ,\
\alpha=1,\ldots,M.
\end{eqnarray}
For the ground state and the excitations we are interested in we only
need to consider real roots of \fr{t1EG}, \fr{t2EG}. For these
the ``quantum numbers'' $I_j$, $J_\a$ fulfil the restrictions
\be
I_j\ {\rm is}\ \bigg\{\begin{array}{l l}
{\rm integer} &\text{if}\ M\ \text{is even}\\
{\rm half-odd\ integer} &\text{if}\ M\ \text{is odd},\\
\end{array}
\label{int-hoi1EG}
\end{equation}
\be
J_\a\ {\rm is}\ \bigg\{\begin{array}{l l}
{\rm integer} &\text{if}\ N-M\ \text{is odd}\\
{\rm half-odd\ integer} &\text{if}\ N-M\ \text{is even},\\
\end{array}
\label{int-hoi2EG}
\end{equation}
The energy and momentum of such Bethe ansatz states are
\be
E=2BM+\sum_{j=1}^N\left[k_j^2-\mu-B\right] ,\qquad
P=\sum_{j=1}^N k_j\equiv\frac{2\pi}{L}\left[
\sum_{j=1}^NI_j+\sum_{\alpha=1}^MJ_\alpha\right].
\label{EPEG}
\ee
\subsection{Ground State and Excitations in the thermodynamic limit}
In the thermodynamic limit the ground state is described by the
solution of the integral equations
\bea
\rho_{c,0}(k)&=&\frac{1}{2\pi}+\int_{-A}^Ad\l
\ \K_{cs}(k-\l)\ \rho_{s,0}(\l)\ ,\nn
\rho_{s,0}(\l)&=&\int_{-Q}^Qdk\ K_{sc}(\l-k)\ \rho_{c,0}(k)
+\int_{-A}^Ad\l' \ \K_{ss}(\l-\l')\ \rho_{s,0}(\l')\ ,
\label{rhocsEG}
\eea
where we have defined integral kernels as
\bea
\K_{cc}(k,k')&=&0\ ,\qquad\qquad\quad\qquad \K_{cs}(k,\l)=a_1(k-\l)\ ,\nn
\K_{sc}(\Lambda,k)&=&a_1(k-\l)\ ,\qquad
\K_{ss}(\l,\l')=-a_2(\l-\l').
\label{kernelEG}
\eea
Density and magnetization per site are given by $n_c$ and $\frac{n_c}{2}-n_s$
respectively, where
\bea
n_c&=&\int_{-Q}^Qdk\ \rho_{c,0}(k) ,\qquad
n_s=\int_{-A}^Ad\l\ \rho_{s,0}(\l) .
\eea
The dressed energies of elementary charge and spin
excitations in the thermodynamic limit at finite density and
magnetization are given in terms of the solutions to the coupled
integral equations 
\bea
\eps_c(k)&=&k^2-\mu-B+\int_{-A}^Ad\l\ \K_{cs}(k-\l)\ \eps_s(\l)\ ,\nn
\eps_s(\l)&=&2B+\int_{-Q}^Qdk\ \K_{sc}(\l-k)\ \eps_c(k)
+\int_{-A}^Ad\l'\ \K_{ss}(\l-\l')\ \eps_s(\l')\ ,
\label{dressedenEG}
\eea
where the integration boundaries are fixed by the requirements
\be
\eps_c(\pm Q)=0\ ,\qquad \eps_s(\pm A)=0.
\ee
The holon-spinon excitation is constructed in complete analogy with
the Hubbard model. In the thermodynamic limit its energy and momentum
are expressed as 
\be
E_{hs}=-\eps_c(k^h)-\eps_s(\l^h)\ ,\quad
P_{hs}=-p_c(k^h)-p_s(\l^h)\pm\pi n_c.
\ee

\subsection{Finite-Size Corrections for Threshold Exitations at
  $|P_{hs}|<k_F$} 
Following through the same steps as for the Hubbard model we obtain
\bea
E&=&Le_{\rm GS}(\{X^\a\})-\eps_s(\l^h)\nn
&&-\frac{\pi}{6L}(v_s+v_c)
+\frac{2\pi}{L}\left[
\frac{1}{4}\Delta \tilde{N}_\gamma (Z^T)^{-1}_{\gamma\a}v_\a Z^{-1}_{\a\beta}
\Delta \tilde{N}_\beta+
 \tilde{D}_\gamma Z_{\gamma\a}v_\a Z^T_{\a\beta}
 \tilde{D}_\beta\right]-\frac{1}{L}\eps'_s(\l^h)\delta\l^h+o(L^{-1})\ .
\label{FSEEG}
\eea
Here $e_{\rm GS}(\{X^\a\})$ is the ground state energy per site in the
thermodynamic limit (and we again use notations where $X_c=Q$ and
$X_s=A$), $-\eps_s(\l^h)$ is the ${\cal O}(1)$ contribution of the
spinon excitation in the thermodynamic limit and
$v_s$ and $v_c$ are the velocities of the gapless elementary spin
and charge excitations. They are given in terms of the solutions of
the integral equations \fr{rhocsEG}, \fr{dressedenEG} by
\bea
v_c&=&\frac{\eps'_c(Q)}{2\pi\rho_{c,0}(Q)}\ ,\qquad
v_s=\frac{\eps'_s(A)}{2\pi\rho_{s,0}(A)}.
\eea
The dressed charge matrix $Z$ in \fr{FSEEG} is defined as
\be
{\bf Z}=
\begin{pmatrix}
\xi_{cc}(Q) & \xi_{cs}(A)\cr
\xi_{sc}(Q) & \xi_{ss}(A)
\end{pmatrix}\ ,
\label{ZEG}
\ee
where $\xi_{\a\b}$ fulfil the set of coupled integral equations
\be
\xi_{\a\b}(z_\b)=\delta_{\a\b}+\sum_{\g=c,s}\int_{-X_\g}^{X_\g}dz_\g\
\xi_{\a\g}(z_\g)\ {\cal K}_{\g\b}(z_\g-z_\b)\ .
\ee
The quantities $\Delta \tilde{N}_\a$ and $\tilde{D}_\a$ are defined
as in the case of the Hubbard model 
\bea
\Delta \tilde{N}_\a&=&\Delta N_\a-N^{\rm imp}_\a\ ,\qquad
\tilde{D}_\a= D_\a-D^{\rm imp}_\a\ ,
\eea
where now
\bea
N_c^{\rm imp}&=&\int_{-Q}^Qdk\ \rho_{c,1}(k)\ ,\qquad
N_s^{\rm imp}=\int_{-A}^Ad\l\ \rho_{s,1}(\l)\ ,\\
2D_c^{\rm imp}&=&\int_{-\infty}^{-Q}-\int_Q^\infty dk\ \rho_{c,1}(k)\ ,\qquad
2D_s^{\rm imp}=\int_{-\infty}^{-A}-\int_A^\infty d\l\ \rho_{s,1}(\l)\ .
\eea
Note that these are different from what we had for the Hubbard model.
The root densities $\rho_{c,1}$ and $\rho_{s,1}$ fulfil the coupled
integral equations
\bea
\rho_{c,1}(k)&=&-a_1(k-\l^h)+\int_{-A}^Ad\l\ \K_{cs}(k-\l)\ \rho_{s,1}(\l)\ ,\\
\rho_{s,1}(\l)&=&a_2(\l-\l^h)+\int_{-Q}^Qdk\ \K_{sc}(\l-k)\ \rho_{c,1}(k)
+\int_{-A}^Ad\l'\ \K_{ss}(\l-\l')\ \rho_{s,1}(\l')\ .
\eea
Finally, $\eps'_s(\l)$ is the derivative of the dressed energy
\fr{dressedenEG} and $\delta \l^h/L$ is the shift in the hole rapidity
due to the finite volume quantization conditions. It is obtained from
a set of equations completely analogous to \fr{dlh1}, \fr{dlh2} and
\fr{dlh3}.

\subsection{Simplification for zero Magnetic Field}
In the absence of a magnetic field we have
$A=\infty$, which again allows us to simplify all expressions. 
The integral equations for the dressed energies can be written in the
form 
\bea
\eps_c(k)&=&k^2-\mu+\int_{-Q}^Qdk^\prime \
R(k^\prime - k) \ \eps_c(k^\prime)\ ,\fnn
\epsilon_s(\Lambda)&=&\int_{-Q}^Qdk\ s(\l-k)\ \eps_c(k)\ ,
\label{EGdresseden0}
\eea
where the integration boundary $Q$ is fixed as a function of the
chemical potential $\mu$ by the requirement
$\eps_c(\pm Q)=0$.
The dressed charge matrix takes the form \fr{xiH0}, where $\xi=\xi(Q)$
is obtained from the integral equation
\be
\xi(k)=1+\int_{-Q}^Qdk'\ \ R(k-k')\ \xi(k').
\ee
The expression for the finite-size energy simplifies to
\bea
E&=&Le_{\rm GS}(\{X^\a\})-\eps_s(\l^h)
-\frac{\pi}{6L}(v_s+v_c)-\frac{1}{L}\eps'_s(\l^h)\delta\l^h\nn
&+&\frac{2\pi  v_c}{L}
\left[\frac{(\Delta N_c-N_c^{\rm imp})^2}{4\xi^2}
+\xi^2\left({D}_c-D^{\rm imp}_c+\frac{{D}_s}{2}\right)^2\right]
+\frac{2\pi
  v_s}{L}
\left[\frac{\left(\Delta {N}_s-\frac{1}{2}\Delta {N}_c-\frac{1}{2}\right)^2}{2}
+\frac{D_s^2}{2}
\right]+o(L^{-1}).
\eea
The expressions for $N^{\rm imp}_\a$ and $D^{\rm imp}_\a$ become
\bea
N^{\rm imp}_c&=&2N_s^{\rm imp}-1=\int_{-Q}^Qdk \rho_{c,1}(k)\ ,\\
2D^{\rm imp}_c&=&
\int_{Q}^{\infty}dk\left[ \rho_{c,1}(-k)- \rho_{c,1}(k)\right],\qquad
D^{\rm imp}_s=0\ ,
\label{NDimpEG}
\eea
where $\rho_{c,1}$ is the solution to the integral equation
\bea
\rho_{c,1}(k)=-\frac{1}{4u\cosh\left(\frac{\pi(\l^h-k)}{2u}\right)}
+\int_{-Q}^Q dk'\ R(k-k')\ \rho_{c,1}(k').
\eea
Here $R(x)$ is given by \fr{rofx}. We note that our expression for
$N_c^{\rm   imp}$ is related to the dressed energies \fr{dressedenEG} by
\be
N_c^{\rm imp}=\frac{\partial\eps_s(\Lambda^h)}{\partial\mu}.
\ee

For the holon-spinon excitation we have
\bea
\Delta N_c&=&N_c-N_{\rm GS}=(N_{\rm GS}-1)-N_{\rm GS}=-1\ ,\nn
\Delta N_s&=&N_s-\frac{N_{\rm GS}}{2}=0.
\eea
Like fo the Hubbard model the corrections in the low-energy spin
sector vanish, which in terms of the field theory description implies
that the interaction of the high-energy spinon with the low-energy
spin sector is irrelevant. 
\subsection{Threshold Exponent}
As for the Hubbard model the spectral function exhibits a threshold
singularity of the form
\be
A(\omega,P)\propto \left(\omega-\eps_s(P)\right)^{-\mu_{0,-}}\ ,
\ee
where the exponent $\mu_{0,-}$ is expressed both in terms of the phase-shifts
$\delta_{\pm,c}$ arising in the field theory treatment \fr{udagger}
and in terms of the quantities $N_c^{\rm imp}$, $D_c^{\rm imp}$ in \fr{NDimpEG}
\bea
\mu_{0,-}&=&1-\frac{1}{2}\left[-\sqrt{\frac{K_c}{2}}
+\frac{\delta_{+,c}+\delta_{-,c}}{2\pi}\right]^2
-\frac{1}{2}\left[\frac{1}{\sqrt{2K_c}}
-\frac{\delta_{+,c}-\delta_{-,c}}{2\pi}\right]^2\nn
&=&1-K_c \Big(2D_c^{\rm imp}\Big)^2-\frac{1}{4K_c}\Big(1+N_c^{\rm imp}\Big)^2.
\label{exponent2}
\eea
In Ref.\onlinecite{SIG} the phase-shifts $\delta_{\pm,c}$ were
expressed in terms of properties of the excitation spectrum. We have
verified the relations given there by numerically solving the relevant
integral equations for the Yang-Gaudin model.
\section{Summary and Conclusions}
In this work we have determined the ${\cal O}(L^{-1})$ corrections to
energies of excited states in the Hubbard and Yang-Gaudin models
for the case where in addition to any finite number of low-energy
excitations, high-energy excitations are present as well. 
This extends the work of Woynarovich \cite{FSspectrum}, which
considered exclusively low-lying excited states.
We have focussed on the case of a single high-energy holon or spinon,
but the method is easily extended to other cases. There are several
contributions to the ${\cal O}(L^{-1})$ energy corrections. One of
these arises from the quantization of the momentum for the high-energy
particle in the finite volume.  More interestingly, we find that the
presence of a high-energy particle leads to a modification of the
low-energy part of the spectrum. This effect can be understood by a
mapping to a model of a mobile impurity coupled to the spin-charge
separated Luttinger liquid that describes the low-energy degrees of
freedom \cite{SIG}. By matching the finite-size spectrum of the mobile
impurity model to the one obtained from the exact solution, we have
obtained explicit results for threshold singularities in the spectral
function $A(\omega, q)$. In the momentum range $|P|<k_F$ the negative
frequency part exhibits a singularity at $\omega=\eps_s(P)$ of the form
\be
A(\omega,P)\propto \Big(\omega-\eps_s(P)\Big)^{-\mu_{0,-}}\ ,
\ee
while the positive frequency part vanishes in a characteristic
power-law fashion above $\omega=-\eps_s(P)$ as
\be
A(\omega,P)\propto \Big(\omega+\eps_s(P)\Big)^{1-\mu_{0,-}}\ .
\ee
Expression \fr{exponent} for $\mu_{0,-}$ is the main result of this
work. The resulting spectral function is depicted in Fig.\ref{fig:aoq}. 

There have been a number of previous studies of the single-particle
spectral function of the one dimensional Hubbard model below
half-filling. Ref.~\onlinecite{jeckel} reports dynamical density
matrix renormalization group results for density $n_c=0.6$ and
$U/t=4.9$. We found that the threshold exponent \fr{exponent} is
only in partial agreement with Ref.~\onlinecite{jeckel}. A possible
explanation is that smaller values for the imaginary part of the
energy are required in the DMRG computation in order to extract the
singularity reliably. It would be interesting to test this conjecture.
In the large-$u$ limit the single-particle spectral function has been
computed \cite{inftyU} by exploiting the factorization of the wave
function into spin and charge parts \cite{ogata}. The behaviour
obtained by this method was reported in Ref.~\onlinecite{carmelo} and
agrees with \fr{expinf}. Finally, singularity exponents obtained by
completely different methods have been reported in
Ref.~\onlinecite{carmelo}. We have checked that the numerical results
for the exponent $\mu_{0,-}$ in the range $|P|<k_F$, density
$n_c=0.59$ and several values of $u$ ($u=0.25,\ 1.225,\ 2.5$) are in
agreement with ours. It would be interesting to demonstrate the
equivalence of the expressions for the exponents of
Ref.~\onlinecite{carmelo} and our results analytically.

\acknowledgments
The work was supported by the EPSRC under grant EP/D050952/1.
I am grateful to L. Glazman, A. Imambekov, E. Jeckelmann and
A.M. Tsvelik for useful discussions.
\appendix
\section{Zero Field Limit}
\label{app:H0}
In this Appendix we consider the zero magnetic field limit for the
quantity $D_s^{\rm imp}$ \fr{Dimp}. By definition we have
\be
D_s^{\rm imp}=\int_{A}^\infty d\l\left[ \rho_{s,1}(-\l)-\rho_{s,1}(\l)\right]=
-\int_0^\infty d\l\ f(\l),
\ee
where 
\be
f(\l)=\rho_{s,1}(\l+A)-\rho_{s,1}(-\l-A).
\ee
After Fourier transforming the integral equation for $\rho_{s,1}$ we
obtain the following set of equations for $f(\l)$ and
$\rho_{c,-}(k)=\rho_{c,1}(k)-\rho_{c,1}(-k)$
\bea
f(\l)&=&-R(\l+A+\l^h)+R(\l+A-\l^h)+\int_{-Q}^Qdk\ s(\l+A-\sin
k)\ \rho_{c,-}(k)\label{f}\\
&&+\int_0^\infty d\l'\left[R(\l-\l')-R(\l+\l'+2A)\right]\ f(\l')\ ,\nn
\rho_{c,-}(k)&=&-\cos(k)\left[s(\l^h-\sin k)-s(\l^h+\sin k)\right]
+\cos(k)\int_{-Q}^Qdk'\ R(\sin k-\sin k')\ \rho_{c,-}(k')\nn
&&-\cos(k)\int_0^\infty d\l\ [s(\l+A-\sin k)-s(\l+A+\sin k)]\ f(\l).
\eea
We now observe that for large $A$ and $|\l^h|\ll \l+A$ the driving term in \fr{f} is small
\be
R(\l+A+\l^h)-R(\l+A-\l^h)\approx-\frac{2u}{\pi}\frac{\l^h}{(\l+A)^3}.
\ee
Iterating the integral equations in this limit then shows
that $D_s^{\rm imp}$ vanishes when $A$ tends to $\infty$.
\section{Finite-Size Corrections for a high-energy holon excitation}
\label{sec:thres3}
Our starting point are the Bethe ansatz equations \fr{BAE} for the
holon-spinon excitation where the spinon sits at the Fermi momentum of
the $\l_\a$'s. The hole in the distribution of $k_j$'s is denoted by
$k^h$ and the correspoinding integer in the logarithmic form of the
Bethe Ansatz equations \fr{BAE} by $I^h$. 
Expressing the Bethe Ansatz equations in terms of counting
functions \fr{CF} we have
\be
z_c(k_j)=\frac{2\pi I_j}{L}\ ,\qquad
z_s(\l_\a)=\frac{2\pi J_\a}{L}\ ,
\ee
where the integers $I_j$ and $J_\a$ are given by
\bea
J_\a&=&-\frac{M_{\rm GS}}{2}+\frac{1}{2}+\a\ ,\quad \a=1,\ldots,M_{\rm
  GS}-1\ ,\nn
I_j&=&
\begin{cases}
-\frac{N_{\rm GS}}{2}+j & \text{if}\ 1\leq j<\frac{N_{\rm GS}}{2}+I^h\cr
-\frac{N_{\rm GS}}{2}+j+1 & \text{if}\ \frac{N_{\rm GS}}{2}+I^h\leq
j<N_{\rm GS}\cr
\end{cases}.
\eea

\subsection{Finite-Size Corrections}

As before we turn these into integral equations by means of the
Euler-Maclaurin sum formula \fr{EMcL}. This results in
\bea
z_c(k)&=&k+\int_{A_-}^{A_+}d\l\ \rho_s(\l)\ 
\th\Bigl(\frac{\sin k-\l}{u}\Bigr)
+\frac{1}{24L^2}\left[
\frac{a_1(\sin k-A_+)}{\rho_s(A_+)}
-\frac{a_1(\sin k-A_-)}{\rho_s(A_-)}\right]+o(L^{-2}),\\
z_s(\l)&=&\int_{Q_-}^{Q_+}dk\
\th\Bigl(\frac{\l-\sin k}{u}\Bigr)\rho_c(k)
-\frac{1}{L}
\th\Bigl(\frac{\l-\sin(k^h_L)}{u}\Bigr)
-\int_{A_-}^{A_+}d\l'\ \rho_s(\l')\ 
\th\Bigl(\frac{\l-\l'}{2u}\Bigr)\nn
&+&\frac{1}{24L^2}\left[
\frac{a_1(\l-\sin Q_+)\cos Q_+}{\rho_c(Q_+)}
-\frac{a_1(\l-\sin Q_-)\cos Q_-}{\rho_c(Q_-)}
-\frac{a_2(\l-A_+)}{\rho_s(A_+)}
+\frac{a_2(\l-A_-)}{\rho_s(A_-)}
\right]+o(L^{-2}),
\label{count2}
\eea
where $a_n(x)$ is given in \fr{an}, the root densities $\rho_{c,s}$
are given in terms of the counting functions by \fr{bcs}, and the
integration boundaries are fixed by 
\be
z_c(Q_\pm)=\frac{2\pi I_{\pm}}{L}\ ,\qquad
z_s(A_\pm)=\frac{2\pi J_{\pm}}{L}.
\ee
Here
\bea
I_\pm=\pm\frac{N_{\rm GS}}{2}+\frac{1}{2}\ ,\qquad
J_+=\frac{M_{\rm GS}}{2}\ ,\quad
J_-=-\frac{M_{\rm GS}}{2}+1.
\eea
The equation fixing the position of the hole is
\be
z_c(k^h_L)=\frac{2\pi I^h}{L}=\ \text{fixed}.
\ee
Here our notation makes the $L$-dependence of the rapidity of the hole
explicit. Following through the same steps as in \ref{sec:thres1}
we find that the finite-size energy is expressed as
\bea
E&=&Le_{\rm GS}(\{X^\a\})-\eps_c(k^h)\nn
&&+\frac{1}{L}\left\{
-\frac{\pi}{6}(v_s+v_c)
+2\pi\left[
\frac{1}{4}\Delta \tilde{N}_\gamma (Z^T)^{-1}_{\gamma\a}v_\a Z^{-1}_{\a\beta}
\Delta \tilde{N}_\beta+
 \tilde{D}_\gamma Z_{\gamma\a}v_\a Z^T_{\a\beta}
 \tilde{D}_\beta\right]-\eps'_c(k^h)\delta k^h\right\},
\label{FSE_2}
\eea
where $e_{\rm GS}(\{X^\a\})$ and $-\eps_c(k^h)$ are respectively the
ground state energy per site and the dressed energy of the holon
in the thermodynamic limit, $Z_{\a\b}$ are the elements of the dressed
charge matrix \fr{Z} and $\Delta \tilde{N}_\a$, $\Delta
\tilde{D}_\a$ are defined by \fr{tildeND}, where
\bea
N_c^{\rm imp}&=&\int_{-Q}^Q dk\ \rho_{c,1}(k)\ ,\qquad
N_s^{\rm imp}=\int_{-A}^A dk\ \rho_{s,1}(\l)\ ,\nn
2D_s^{\rm imp}&=&\int_{-\infty}^{-A}d\l \rho_{s,1}(\l)
-\int_{A}^{\infty}d\l \rho_{s,1}(\l)\ ,\nn
2D_c^{\rm imp}&=&\int_{-\pi}^{-Q}dk\ \rho_{c,1}(k)
-\int_{Q}^{\pi}dk\ \rho_{c,1}(k)-\frac{1}{\pi}\int_{-A}^{A}d\l\
\th\left(\frac{\l}{u}\right)\ \rho_{s,1}(\l).
\label{Dimp2}
\eea
Here the root densities $\rho_{\a,1}$ fulfil the coupled integral equations
\bea
\rho_{c,1}(k)&=&\cos(k)\int_{-A}^Ad\l
\ a_1(\l-\sin(k))\ \rho_{s,1}(\l)\ ,\nn
\rho_{s,1}(k)&=&-a_1(\l-\sin(k^h))+\int_{-Q}^Qdk\ a_1(\l-\sin k)\ \rho_{c,1}(k)-
\int_{-A}^Ad\l'
\ a_2(\l-\l')\ \rho_{s,1}(\l')\ .
\eea
The quantum numbers $D_\a$, $\Delta N_\a$ are
\be
 D_c= D_s=\frac{1}{2}\ ,\qquad
\Delta N_c=0\ ,\quad \Delta N_s=-1.
\ee 
Like for the spinon threshold it is possible to express $N_{c,s}^{\rm
  imp}$ in terms of the dressed energies. We find that
\be
N_c^{\rm imp}=1+\frac{\partial\eps_c(k^h)}{\partial\mu}\ ,\qquad
N_s^{\rm imp}=-\frac{1}{2}\left[
\frac{\partial\eps_c(k^h)}{\partial B}-
\frac{\partial\eps_c(k^h)}{\partial\mu}\right] .
\ee

\subsection{Simplification for zero Magnetic Field}
In zero magnetic field the expression for the energy simplifies to
\bea
E&=&Le_{\rm GS}(\{X^\a\})-\eps_c(k^h)
-\frac{\pi}{6L}(v_s+v_c)-\frac{1}{L}\eps_c'(k^h)\delta k^h\nn
&+&\frac{2\pi
  v_c}{L}
\left[\frac{(\Delta N_c-N_c^{\rm imp})^2}{4\xi^2}
+\xi^2\left({D}_c-D^{\rm imp}_c+\frac{{D}_s}{2}\right)^2\right]
+\frac{2\pi
  v_s}{L}
\left[\frac{\left(\Delta {N}_s-\frac{\Delta {N}_c}{2}+\frac{1}{2}\right)^2}{2}
+\frac{{D}_s^2}{2}\right], 
\eea
where
\bea
N^{\rm imp}_c&=&2N_s^{\rm imp}+1=\int_{-Q}^Qdk \rho_{c,1}(k)\ ,\qquad
D^{\rm imp}_s=0\ ,\\
2D^{\rm imp}_c&=&
\int_{Q}^{\pi}dk\left[ \rho_{c,1}(-k)- \rho_{c,1}(k)\right]
-\frac{1}{\pi}\int_{-Q}^Qdk\left[\rho_{c,1}(k)-\delta(k-k^h)\right]
i\ln\left[
\frac{\Gamma\left(\frac{1}{2}+i\frac{\sin k}{4u}\right)}
{\Gamma\left(\frac{1}{2}-i\frac{\sin k}{4u}\right)}
\frac{\Gamma\left(1-i\frac{\sin k}{4u}\right)}
{\Gamma\left(1+i\frac{\sin k}{4u}\right)}\right].
\eea
Here $\rho_{c,1}$ is the solution to the integral equation
\bea
\rho_{c,1}(k)=-\cos(k)R(\sin(k)-\sin(k^h))
+\cos(k)\int_{-Q}^Q dk'\ R(\sin k-\sin k')\ \rho_{c,1}(k').
\eea


\end{document}